\documentclass[manuscript,noblind]{geophysics}

\usepackage{amssymb}
\usepackage{amsmath}
\usepackage{empheq}
\usepackage{amscd}
\usepackage{graphicx}
\usepackage{graphics}
\usepackage[noadjust]{cite}
\usepackage{amsthm}
\usepackage{caption}
\usepackage{multirow}
\usepackage{array}
\usepackage{rotating}
\usepackage{algorithmic}
\usepackage{algorithm}
\usepackage{multicol}

\usepackage{indentfirst}
\usepackage{tikz}
\usepackage{calc}
\usepackage{epsfig}
\usepackage{natbib}
\usepackage[makeroom]{cancel}
\usepackage{multicol}
\usepackage{csquotes}
\usepackage{enumitem}

\usepackage[utf8]{inputenc}
\usepackage{optidef}
\usepackage{hyperref}
\hypersetup{colorlinks,linkcolor={blue},citecolor={blue},urlcolor={red}}

\DeclareMathOperator*{\argmin}{argmin}

\begin{document}
\title{A practical implementation of data-space Hessian in the time-domain extended-source full-waveform inversion}

\address{ \footnotemark[1]University Cote d'Azur - CNRS - IRD - OCA, Geoazur, Valbonne, France, email: guogs@geoazur.unice.fr, operto@geoazur.unice.fr, aghamiry@geoazur.unice.fr \\
\footnotemark[2]University of Tehran, Institute of Geophysics, Tehran, Iran, email: agholami@ut.ac.ir
\footnotemark[3]Institute of Geophysics, Polish Academy of Sciences, Warsaw, Poland, email: agholami@igf.edu.pl 
}
\author{Gaoshan Guo \footnotemark[1], St\'ephane Operto \footnotemark[1], Ali Gholami \footnotemark[2],\footnotemark[3] \\
Hossein S. Aghamiry \footnotemark[1] }

\lefthead{Guo et al.}
\righthead{Time-domain extended-source FWI}

\maketitle
\clearpage
\begin{abstract}
Full-waveform inversion (FWI) with extended sources, also called wavefield reconstruction inversion, first computes wavefields with data-driven source extensions, i.e. wave equation errors, such that the simulated data in inaccurate velocity models match the observed counterpart well enough to prevent cycle skipping. Then, the source extensions are minimized to update the model parameters toward the true medium. This two-step workflow is iterated until both data and sources are matched. It was recently shown that the source extensions are the least-squares solutions of the recorded scattered data (i.e., FWI data residuals) fitting problem. As a result, they are computed by propagating backward in time the deblurred FWI data residuals, where the deblurring operator is the inverse of the damped data-domain Hessian of the scattering-source estimation problem. Estimating the deblurred data residuals is the main computational bottleneck of time-domain extended-source FWI. To mitigate this issue, we first estimate them when the inverse of the data-domain Hessians is approximated by matching filters in Fourier and short-time Fourier domains. Second, we refine them with conjugate-gradient iterations when necessary. Computing the matching filters and performing one conjugate-gradient iteration each require two simulations per source. Therefore, it is critical to design some workflows that minimize the footprint of this computational burden. We implement time-domain extended-source FWI with the augmented Lagrangian method. Moreover, we further extend its linear regime with a multiscale frequency continuation approach, which is combined with grid coarsening to mitigate the computational burden and regularize the inversion by reparametrization. Finally, we use total-variation regularization to deal with large-contrast reconstruction. We present synthetic case studies (Marmousi II and 2004 BP salt models) where different inversion workflows carried out with data-domain Hessians of variable accuracy were assessed with the aim at converging toward accurate solutions while minimizing computational cost.
\end{abstract}
\graphicspath{{"./figures/"}}

\section{Introduction}

\noindent Full waveform inversion (FWI) has become the baseline method for determining high-resolution subsurface models. However, the objective function of conventional FWI, namely the least-squares ($\ell{2}$) norm of waveform differences, is highly multimodal as manifested by cycle-skipping that drives FWI towards spurious solutions when simulated traveltimes don't match recorded counterpart with an error lower than half the period \citep{Gauthier_1986_TDN}. During the last decade, several methods have been proposed to face this long-term problem by starting from an accurate initial model obtained by reflection tomography or stereo-tomography \citep{Sambolian_2019_PST}, applying data-driven hierarchical schemes \citep{Shipp_2002_TDF,Gorszczyk_2017_TRW}, proposing new distances between recorded and simulated data \citep[e.g.][]{Ma_2013_WRT,Baek_2014_RLS,Metivier_2015_MTM, Warner_2016_AWI, Yang_2017_AOT}, and adding non-physical degrees of freedom in the forward problem to extend the search space of the inversion \citep{Symes_2008_MVA,VanLeeuwen_2013_MLM, Biondi_2014_SIF,vanLeeuwen_2016_PMP,Wang_2016_FIR, Huang_2018_SEW}. Among the extended-space approaches, some extend the search space in the model domain with time lags or subsurface offsets \citep{Symes_2008_MVA, Biondi_2014_SIF,Barnier_2022_FWI}, while others implement this extension in the source domain. This paper focuses on the second category of extended-space approaches, which gather the so-called extended-source (ES-)FWI \citep{Wang_2016_FIR, Huang_2018_SEW}, the wavefield reconstruction inversion (WRI) method \citep{VanLeeuwen_2013_MLM,vanLeeuwen_2016_PMP} and the contrast-source inversion (CSI) method \citep{Abubakar_2009_FDC}. These methods extend the search space of FWI by allowing for source extension (i.e., wave-equation errors) during each wavefield reconstruction such that simulated data in inaccurate velocity models match the recorded counterpart accurately enough to prevent cycle skipping. Then, the source extensions are minimized to update the model parameters toward the true medium assuming that the reconstructed data-assimilated wavefields are good approximation of the true wavefields. This two-step workflow is iterated until both the data and the source are matched. Broadly speaking, ES-FWI, WRI and CSI rely on the same principle. They mainly differ in the optimization variables, which can involve the wavefield, the source or the contrast source formed by the interaction of the incident wavefield with the scattering object. These extended-space approaches are classically implemented by processing the wave equation as a soft constraint with penalty methods. However, penalty methods suffer from slow convergence, require adaptive tuning of the penalty parameter \citep{Fu_2017_DPM} and may converge toward suboptimal minimizer \citep[][ chapter 17]{Nocedal_2006_NO}. More efficient optimization schemes can be designed with augmented Lagrangian methods as proposed by \citet{Aghamiry_2019_IWR} and \citep{Gholami_2022_EFW} in the frequency domain and in the time domain, respectively. They called their approach iteratively-refined(IR)-WRI.

\noindent WRI/IR-WRI have been widely implemented in the frequency domain where the reconstructed wavefields can be reconstructed with linear algebra methods as the least-squares solution of an overdetermined linear system gathering the wave equation and the observation equation \citep{VanLeeuwen_2013_MLM,vanLeeuwen_2016_PMP}. Conversely, wavefield reconstructions are more problematic in the time domain with explicit time-stepping schemes because the source extension in the right-hand side of the relaxed wave equation depends on the data residuals in the extended space (and hence on the unknown sought wavefields) \citep{Aghamiry_2020_AED}. However, \citet{Gholami_2022_EFW,Operto_2022_FBA} recognized that these source extensions or scattering sources are the least-squares($\ell{2}$) solutions of the underdetermined quadratic scattered-data fitting problem, and hence can be computed by time-reversed modeling of the FWI data residuals (i.e., measured scattered data) that are beforehand deblurred by the inverse of the data-domain Hessian. \citet{Gholami_2022_EFW} solved this least-squares problem with one iteration of a steepest-descent algorithm, which means that the data-domain Hessian is approximated by the identity matrix and the step length is estimated by minimizing the residuals between the observed and predicted data. This approximation is acceptable in some favorable settings (subsurface medium of moderate complexity, sufficiently-accurate starting model) but may fail for complex media and inaccurate starting models.

\noindent In this study, we estimate more accurately the deblurred data residuals by solving iteratively the Gauss-Newton system of the scattered-data fitting problem with a conjugate gradient (CG) solver when a good initial guess is used to start the CG loop. These initial guesses are estimated by applying an approximate inverse of the data-domain Hessian on the FWI data residuals, where the approximate Hessian inverse is built by a 1D and 2D least-squares matching filter designed in the Fourier domain \citep{Liu_2018_ODL} or the short-time Fourier domain with a Gaussian window \citep{Liu_2019_SDL}. We first show the effect of considering the data-domain Hessian on the simulated data in the extended space before showing its impact on the waveform inversion applied to the reconstruction of Marmousi II and the 2004 BP salt model from crude starting models. We also illustrate the key role of sparsity-promoting total-variation (TV) regularization for imaging large contrasts and discuss the effects of surface multiples on the waveform inversion. Finally, we propose a practical workflow combining IR-WRI and classical FWI allowing for accurate subsurface reconstruction while minimizing the computational cost. Lastly, we discuss the limitation of the current study and possible strategies to improve it in future studies.

\section{Theoretical background}

\noindent We first review the first-order acoustic wave equation and its self-adjoint form that are used as modeling engine and as state equation in the forward problem and in the inverse problem, respectively. Then, we review the basic principles of extended source FWI with a special emphasis on the expression and the role of the source extension in this FWI formulation.

\subsection*{State equation}
\noindent We perform seismic modeling with a finite-difference staggered-grid method \citep{Virieux_1984_SWP}  applied on the 2D velocity-stress acoustic wave equation with a constant density,
\begin{equation}
\begin{cases}
\partial_t v_x(\mathbf{x},t) = \partial_x p (\mathbf{x},t), \\
\partial_t v_z(\mathbf{x},t)= \partial_z p (\mathbf{x},t), \\
\partial_t p(\mathbf{x},t)= c^2(x) \left( \partial_x v_x (x,t) +  \partial_z v_z (\mathbf{x},t) + s(\mathbf{x},t) \right),
\end{cases}
\label{1steq}
\end{equation}
where $v_x(\mathbf{x},t) \in \mathbb{R}^{m \times n_t}$, $v_z(\mathbf{x},t)\in \mathbb{R}^{m \times n_t}$ and $p(\mathbf{x},t)\in \mathbb{R}^{m \times n_t}$ are the horizontal and vertical particle velocities and pressure wavefields, respectively, $s(\mathbf{x},t)=s(t) \delta (\mathbf{x}-\mathbf{x}_s )\in \mathbb{R}^{m \times n_t}$ is the point source, $c(\mathbf{x}) \in \mathbb{R}^{m}$ is the subsurface velocity model, and $m$ and $n_t$ are the numbers of spatial and temporal grid points. Extensions to more complex wave physics and higher spatial dimensions do not raise particular difficulties. \\
The wave equation, equation~\ref{1steq}, can be written in a more compact form as
\begin{equation}
\partial_t \mathbf{u}(\mathbf{x},t)=\mathbf{M}(\mathbf{m}(\mathbf{x})) \left( \mathbf{D} \mathbf{u}(\mathbf{x},t) + \mathbf{b}(\mathbf{x},t) \right),
\end{equation}
where 
\begin{equation}
\mathbf{M}(\mathbf{m}(\mathbf{x}))=\begin{pmatrix}
1 & 0 & 0 \\
0 & 1 & 0 \\
0 & 0 & c^2(\mathbf{x})
\end{pmatrix},~
\mathbf{D}=\begin{pmatrix}
0 & 0 & \partial_x \\
0 & 0 & \partial_z \\
\partial_x  & \partial_z & 0
\end{pmatrix},
\mathbf{b}(\mathbf{x},t)=\begin{pmatrix}
0  \\
0  \\
s(\mathbf{x},t)
\end{pmatrix},
\mathbf{u}(\mathbf{x},t)=\begin{pmatrix}
v_x(\mathbf{x},t) \\
v_z(\mathbf{x},t) \\
p(\mathbf{x},t)
\end{pmatrix}
\end{equation}
\noindent with initial conditions: $\mathbf{u}(\mathbf{x}, 0) = 0$ and $\partial_t \mathbf{u}(\mathbf{x}, 0) = 0$.

\noindent To manipulate self-adjoint operators in the inverse problem, we move the subsurface parameters in the left-hand side \citep[e.g., ][]{Vigh_2014_EFI,Yang_2016_SFM}:
\begin{equation}
\mathbf{C}(\mathbf{m}(\mathbf{x})) \partial_t\mathbf{u}(\mathbf{x},t) = \mathbf{Du}(\mathbf{x},t) + \mathbf{b}(\mathbf{x},t),
\label{wave}
\end{equation}
\noindent where
\begin{equation}
\mathbf{C}(\mathbf{m}(\mathbf{x}))=\mathbf{M}(\mathbf{m}(\mathbf{x}))^{-1}=\begin{pmatrix}
1 & 0 & 0 \\
0 & 1 & 0 \\
0 & 0 & 1/c^2(\mathbf{x})
\end{pmatrix}.
\end{equation}
\noindent For sake of compact notation, we recast the state equation \ref{wave} in matrix form as
\begin{equation}
\mathbf{A(m)u}=\mathbf{b},
\end{equation}
\noindent where $\mathbf{A(m)}=\mathbf{C(m)} \partial_t - \mathbf{D}$. The expression of the adjoint operator $\mathbf{A}^T(\mathbf{m})$, which is needed in the next section, is developed in the Appendix A.

\subsection{WRI or extended-source FWI}
\noindent WRI extends the search-space of FWI by processing the wave equation as a soft constraint with a penalty method to fit the data with an arbitrary accuracy from a given background model, hence avoiding cycle skipping \citep{VanLeeuwen_2013_MLM,vanLeeuwen_2016_PMP,Wang_2016_FIR,Huang_2018_SEW}. \\
The multivariate objective function formulated as a penalty function reads
\begin{equation}
\min_{\mathbf{u}_s,\mathbf{m}} \mathcal{P}_{\mu}(\mathbf{u}_s,\mathbf{m}) =  \frac{1}{2} \sum_{s=1}^{n_s} \| \mathbf{P}_s \mathbf{u}_s -\mathbf{d}_s^* \|_2^2 + \frac{\mu}{2} \sum_{s=1}^{n_s} \| \mathbf{A}(\mathbf{m})\mathbf{u}_s - \mathbf{b}_s^* \|_2^2,
\label{eqwri}
\end{equation}
\noindent where  $n_s$ is the number of sources, $\mathbf{u}_s \in \mathbb{R}^{(3 \times m \times n_t)}$ is the wavefield triggered by the source $s$, $\mathbf{d}_s^* \in \mathbb{R}^{n_r \times n_t}$ are the pressure data for source $s$ (the subscript $*$ refers to measured or true quantities), $\mathbf{b}_s^*$ is the source $s$, $\mathbf{P} \in \mathbb{R}^{(n_r \times n_t) \times (3 \times m \times n_t)}$ is the linear observation operator sampling the pressure wavefield at receiver positions, $n_r$ is the number of receivers (we assume a fixed-spread acquisition but our approach applies to other acquisitions), $m$ is the number of grids, $n_t$ is the number of temporal sampling and the scalar $\mu \in \mathbb{R}_{+}$ is the penalty parameter that controls the amount of relaxation of the wave equation and hence the data fitting. The multivariate problem for $\mathbf{u}_s$ and $\mathbf{m}$, equation~\ref{eqwri}, can be solved with alternating directions or by variable projection taking advantage that the subproblems for $\mathbf{u}_s$ are linear \citep{vanLeeuwen_2016_PMP}. 


\subsubsection{Wavefield reconstruction in the time domain}
\noindent The solution $\mathbf{u}^e$ of the overdetermined $\mathbf{u}$-subproblem satisfies the following normal equation:
\begin{equation}
\left(\mu \mathbf{A}^T(\mathbf{m}) \mathbf{A}(\mathbf{m})+ \mathbf{P}^T \mathbf{P} \right) \mathbf{u}^e = \mu \mathbf{A}^T(\mathbf{m}) \mathbf{b}^* + \mathbf{P}^T \mathbf{d}^*.
\label{eqnormal}
\end{equation}
In the above equation, the superscript $^e$ refers to the extended space as opposed to the reduced space (denoted by superscript $^r$ in the following) where the wave equation is strictly satisfied.
We also drop the subscript $s$ for sake of compact notation. However, it should be remembered that equation~\ref{eqnormal} needs to be solved for all $s$ as is the case for the FWI forward equation.
While the normal system, equation~\ref{eqnormal}, can be solved in the frequency domain with linear algebra methods, it is impractical for time-domain modeling with explicit time-stepping schemes \citep{Aghamiry_2019_AEW,Gholami_2022_EFW}. \\
To overcome this issue, we recast the wavefield reconstruction problem as a scattering-source estimation problem. Let's first rewrite $\mathbf{u}^e$ as the minimizer of the monovariate objective function considering $\mathbf{m}$ fixed:
\begin{equation}
\mathbf{u}^e= \text{arg} \min_{\mathbf{u}} \mathcal{P}_{\mu}(\mathbf{u})|_{\mathbf{m}} =  \frac{1}{2} \| \mathbf{P} \mathbf{u} -\mathbf{d}^* \|_2^2 + \frac{\mu}{2} \| \mathbf{A}(\mathbf{m})\mathbf{u} - \mathbf{b}^* \|_2^2,
\label{eque}
\end{equation}
\noindent We perform the change of variable $\bold{u} \rightarrow \delta \mathbf{b}$ in the above equation
\begin{equation}
\delta \mathbf{b} = \mathbf{A} \mathbf{u} - \mathbf{b}^*,
\end{equation}
where $\delta \mathbf{b}$ is the wave equation error or source extension generated by the wave-equation relaxation. 
The minimizer of the  $\delta \mathbf{b}$-subproblem is given by

\begin{equation}
\delta \mathbf{b}^e = \text{arg} \min_{\delta \mathbf{b}} \frac{1}{2} \| \mathbf{S}(\mathbf{m}) \delta \mathbf{b} -\delta \mathbf{d}^* (\mathbf{m})\|_2^2 + \frac{\mu}{2} \| \delta \mathbf{b} \|_2^2,
\label{eqwridb}
\end{equation}
where $ \mathbf{S}(\mathbf{m})  = \mathbf{P}\mathbf{A}^{-1}(\mathbf{m})$ is the rank-deficient forward-modeling operator and $\delta \mathbf{d}^* (\mathbf{m})=\mathbf{d}^*-\mathbf{P}\mathbf{u}^r(\mathbf{m})=\mathbf{d}^*-\mathbf{S}(\mathbf{m})\mathbf{b}^*$ are the data residuals of classical FWI. \\
Since $\bold{u}$ and $\delta \mathbf{b}$ are linearly related and $\mathbf{A}$ is a full-rank matrix, we have
\begin{equation}
\mathbf{u}^e=\mathbf{u}^r + \delta \bold{u}^e = \mathbf{A}^{-1} \left( \mathbf{b}^* + \delta \mathbf{b}^e \right).
\label{eque1}
\end{equation}
where $\delta \bold{u}^e$ is the scattered field by $\delta \mathbf{b}^e$.
That is, if we can estimate the minimizer of the optimization problem for $\delta \mathbf{b}$, equation~\ref{eqwridb}, then we can compute the minimizer of the optimization problem for $\mathbf{u}^e$, equation~\ref{eque}, by solving the wave equation with an explicit time-stepping scheme and the extended source $\mathbf{b}^* + \delta \mathbf{b}^e$, equation~\ref{eque1}.
The minimizer $\delta \mathbf{b}^e$ of the problem $\ref{eqwridb}$ has a closed-form expression given by
\begin{equation}
\begin{aligned}
\delta \mathbf{b}^e & = \left( \mathbf{S(m)}^T \mathbf{S(m)} + \mu \mathbf{I} \right)^{-1} \mathbf{S(m)}^T \delta \mathbf{d}^*(\mathbf{m}) =  \mathbf{H}_s(\mathbf{m})^{-1} \mathbf{S(m)}^T \delta \mathbf{d}^*(\mathbf{m}) \\
& = \mathbf{S(m)}^T \left( \mathbf{S(m)} \mathbf{S(m)}^T + \mu \mathbf{I} \right)^{-1} \delta \mathbf{d}^*(\mathbf{m}) = \mathbf{S(m)}^T \mathbf{H}_d(\mathbf{m})^{-1} \delta \mathbf{d}^*(\mathbf{m}), \label{eqdbe}
\end{aligned}
\end{equation}
where $\mathbf{H}_s(\mathbf{m})\in \mathbb{R}^{(m \times n_t) \times (m \times n_t)}$ and $\mathbf{H}_d(\mathbf{m}) \in \mathbb{R}^{(n_r \times n_t) \times (n_r \times n_t)}$ denote the source-domain Hessian and the data-domain Hessian, respectively. \\
It is indeed more computationally efficient to implement the data-domain formulation since $n_r << m$. However, accounting for these data-domain Hessians for each source remains the main bottleneck of the time domain formulation of WRI. \\
Substituting the expression of $\delta \mathbf{b}^e$ in equation~\ref{eque1} gives the wave equation satisfied by $\mathbf{u}^e$:
\begin{equation}
\mathbf{A}(\mathbf{m})\mathbf{u}^e = \mathbf{b}^* + \mathbf{S(m)}^T \left( \mathbf{S(m)} \mathbf{S(m)}^T + \mu \mathbf{I} \right)^{-1} \delta \mathbf{d}^*(\mathbf{m}).
\label{eque3}
\end{equation}
This equation shows that $\mathbf{u}^e$ can be computed in four steps: \\
1) Compute the FWI data residuals $\delta \mathbf{d}^*(\mathbf{m})$. \\
(2) Solve the normal system for the deconvolved data residuals $\delta \bold{d}^{d*}$:
\begin{equation}
\left( \mathbf{S(m)} \mathbf{S(m)}^T + \mu \mathbf{I} \right)^{-1} \delta \bold{d}^{d*} = \delta \mathbf{d}^*(\mathbf{m}).
\label{eqnormaldd}
\end{equation}
(3) Solve an adjoint wave equation for $\delta \mathbf{b}^e$ using the weighted data residuals as source:
\begin{equation}
 \mathbf{A(m)}^T \delta \mathbf{b}^e = \bold{P}^T \delta \bold{d}^{d*}.
\end{equation}
(4) Solve the wave equation for $\mathbf{u}^e$ with the extended source $\mathbf{b}^*  + \delta \mathbf{b}^e$, equation~\ref{eque3}. \\
%
%
%
%
%


\subsubsection{Subsurface parameter updating}
\noindent Whatever $\bold{u}$ and $\bold{m}$ are updated in alternating mode or by variable projection, the gradient of $\mathcal{P}_{\mu}(\mathbf{u},\mathbf{m})$ with respect to $\mathbf{m}$ is equal to the gradient of the least-squares norm of wave equation error \citep{vanLeeuwen_2016_PMP}.
\begin{eqnarray}
\nabla_{\mathbf{m}} \mathcal{P}_{\mu}(\mathbf{u},\mathbf{m}) &= & \frac{1}{2} \sum_{s=1}^{n_s} \nabla_{\mathbf{m}} \| \mathbf{A}(\mathbf{m})\mathbf{u}^e_s - \mathbf{b}_s \|_2^2 \label{esfwigrad0} \\
 & = & \sum_{s=1}^{n_s} \left( \frac{\partial \mathbf{A}(\mathbf{m}) \mathbf{u}^e_s}{\partial \mathbf{m}} \right)^T \left( \mathbf{A}(\mathbf{m})\mathbf{u}^e_s - \mathbf{b}_s \right) \label{esfwigrad1} \\
 & = & \sum_{s=1}^{n_s} \left( \frac{\partial \mathbf{A}(\mathbf{m}) \mathbf{u}^e_s}{\partial \mathbf{m}}\right)^T \mathbf{S}^T(\mathbf{m}) \mathbf{H}_{d}^{-1}(\mathbf{m}) \delta \mathbf{d}_s^*(\mathbf{m}) \label{esfwigrad2}
\end{eqnarray}
The gradients of WRI (equation~\ref{esfwigrad2}) and FWI differ in two aspects. The extended wavefields $\mathbf{u}^e$ replace $\mathbf{u}^r$ in the so-called virtual source of the partial derivative data (term in brackets in  equations~\ref{esfwigrad1}-\ref{esfwigrad2}) and the FWI data residuals are weighted by the inverse of the data-domain Hessian in the source of the adjoint equation (right term in equation~\ref{esfwigrad2}). In this framework, the estimated scattering sources $\delta \bold{b}_s^e$ take action in two different places: First, as the adjoint wavefields in the right part of equations~\ref{esfwigrad1}-\ref{esfwigrad2}, and second as the source of the scattered wavefield $\delta \bold{u}^e$ added to the background wavefield to form the data-assimilated wavefield $\bold{u}^e$, equation~\ref{eque1}, in the virtual sources of the partial derivative data (term in brackets in equations~\ref{esfwigrad1}-\ref{esfwigrad2}). Therefore, we can anticipate that the accuracy with which we estimate these scattering sources (i.e., the accuracy with which we account for the effect of the data-domain Hessian) can have a significant impact on the solution of the extended-space waveform inversion. The reader is referred to \citet{Operto_2022_FBA} for a more comprehensive review of the role of these terms in WRI. In this study, we update $\bold{u}$ and $\bold{m}$ in alternating mode. In this case, the Hessian of the $\bold{m}$-subproblem is diagonal and is simply formed by the auto-correlation of the virtual scattering sources, namely the so-called pseudo-Hessian of \citet{Shin_2001_IAP}. The reader is referred to 
\citet{Gholami_2022_EFW} for the expression of the full Hessian when $\bold{m}$ is updated with a variable projection method.


\subsubsection*{From penalty method to augmented Lagrangian method with regularization}

\noindent To converge toward more accurate minimizers while avoiding the tedious tuning of $\mu$,\citet{Aghamiry_2019_IWR} and \citet{Aghamiry_2019_IBC,Gholami_2022_EFW} implemented WRI with the  augmented Lagrangian method leading to the so-called iteratively-refined WRI. The constrained optimization problem is recast as the minimization of a regularization term subject to the observation-equation and wave-equation constraints. This constrained problem is implemented with augmented Lagrangian function, which combines a penalty function to allow for the initial constraint relaxation and a Lagrangian term, the role of which is to help enforcing the constraint at the convergence point through the defect correction action of the Lagrange multipliers \citep[][ chapter 17]{Nocedal_2006_NO}. Therefore, compared to penalty function, the augmented Lagrangian function has two leverages to satisfy the constraint at the convergence point, namely the penalty parameter and the Lagrange multipliers. The former is classically disabled by keeping it fixed hence avoiding its tedious tuning.
In this framework, the regularized objective function becomes:
\begin{eqnarray}
\min_{\mathbf{m},\mathbf{u}} \max_{\boldsymbol{\lambda},\boldsymbol{\nu}} \mathcal{L}_{\mu}(\mathbf{m}, \mathbf{u},\boldsymbol{\lambda},\boldsymbol{\nu})  =  \mathcal{R}(\mathbf{m}) & + & (\frac{1}{2} \| \mathbf{P} \mathbf{u}^e -\mathbf{d}^* \|_2^2 + \frac{\mu}{2} \| \mathbf{A(m)u}^e - \mathbf{b}^* \|_2^2 \nonumber \\
& - & \left\langle \boldsymbol{\lambda}, \mathbf{Pu} - \mathbf{d}^* \right\rangle - \left\langle \boldsymbol{\nu}, \mathbf{A(m)u} - \mathbf{b}^* \right\rangle.
\end{eqnarray}
This represents a saddle point problem where the objective function is minimized with respect to the primal variables $\mathbf{u}$ and $\mathbf{m}$ and maximized with respect to the dual variables (Lagrange multipliers) $\boldsymbol{\lambda}$ and $\boldsymbol{\nu}$.
In the framework of the method of multipliers, the primal and the dual variables are updated in alternating mode. In this case, the dual variables can be updated with basic gradient ascent steps and reduce to the running sum of the constraint errors weighted by the penalty parameter. In this study, we also update the two primal variables $\mathbf{u}$ and $\mathbf{m}$ in alternating mode rather than through variable projection in the framework of the alternating-direction method of multipliers (ADMM).
Compared to the penalty formulation, the only modification consists of adding the running sum of the weighted data residuals from previous iterations to the weighted data residuals of the current iteration in the adjoint source of the $\mathbf{m}$-subproblem  \citep[][ their equations 31, 32 and 36]{Gholami_2022_EFW}. 
In this study, $\mathcal{R}(\mathbf{m})$ is the first-order isotropic total variation of the model $\bold{m}$. We implement this regularization with ADMM to decouple the $\ell{1}$ subproblems from the $\ell{2}$ counterpart and solve the former with computationally-efficient proximal algorithms \citep{Goldstein_2009_SBM,Parikh_2013_PA}. The reader is referred to \citet{Aghamiry_2019_IBC} for more details and Appendix B for short review.


\section{Implementing data-domain Hessian in extended-source FWI }

\noindent The computational bottleneck of time-domain extended-source FWI results from the computation of the source extension $\delta \mathbf{b}^e$, equation~\ref{eqdbe}. This computation first requires the estimation of the source $\delta \mathbf{d}^{d*}$ of the adjoint equation satisfied by $\delta \mathbf{b}^e$. This adjoint source is itself the solution of the normal system involving the data-domain Hessian of the $\delta \mathbf{b}^e$-subproblem, equation~\ref{eqnormaldd}. The accuracy with which this normal system is solved is passed on the accuracy with which the extended source is estimated and hence on the ability of the extended wavefield $\mathbf{u}^e$ to fit the data.
\citet{Gholami_2022_EFW} solve the scattering-source estimation problem with a steepest-descent algorithm, i.e., $\mathbf{H}_d(\mathbf{m}) \approx \gamma \mathbf{I}$, where the step length $\gamma$ is estimated by fitting the observed and the simulated data in a $\ell{2}$ sense source by source. Hereafter this algorithm is referred to as the scalar-fitting (SF) algorithm. SF algorithm limits the performance of the extended-source FWI for complicated and highly-contrasted models when the inversion starts from a rough initial model. In this study, we account for the effect of the data-domain Hessian more accurately. First we approximate its inverse with 1D and 2D matching filters. Then, we use this approximation as a starting guess to solve the source-dependent normal systems for $\delta \mathbf{b}^e$, equation~\ref{eqdbe}, with linear conjugate-gradient (CG) iterations. 

\subsection*{Accounting for data-domain Hessian with CG iteration}
\noindent We solve the normal system for $\delta \bold{d}^{d*}$, equation~\ref{eqnormaldd}, with the CG method. The cost function for CG is defined as
\begin{equation}
f(\delta \mathbf{d}^{d*})=\frac{1}{2} {\delta \mathbf{d}^{d*}}^T \mathbf{H}_d(\mathbf{m}) {\delta \mathbf{d}^{d*}} - {\delta \mathbf{d}^{d*}}^T \delta \mathbf{d}^*
\end{equation}
One CG iteration reads
\begin{equation} \label{CG_iter}
\delta \mathbf{d}^{d*}_{l+1} = \delta \mathbf{d}^{d*}_{{l}} + \alpha_{l} \mathbf{p}_{l},
\end{equation}
where
\begin{equation}
\mathbf{p}_{l} = -\mathbf{r}_{l} + \beta_{l} \mathbf{p}_{{l-1}},
\end{equation}
and $l$ denotes the CG iteration count, $\mathbf{p}_{0}=\mathbf{r}_{0}$, $\mathbf{r}_{l}= \mathbf{H}_d \delta \mathbf{d}^{*d}_{l} - \delta \mathbf{d}^*$, and $\alpha_{l}$ and $\beta_{l}$ are scalars defined in \citet[][ page 119]{Nocedal_2006_NO}. 
The computational cost per CG iteration scales to the cost of one Hessian-vector product. This product requires two wavefield simulations per source, one to propagate backward in time the input vector, one to propagate forward in time the resulting wavefield. 
Due to significant computational overhead generated by the accounting for the Hessian, it is crucial to assess the accuracy with which the normal system needs to be solved at a given IR-WRI iteration and define accordingly an appropriate stopping criterion of iterations to mitigate the number of CG iterations. Classically, the stopping criterion of iteration is defined according the relative backward error $\epsilon_1$,
\begin{equation}
\epsilon_1=\frac{\| \mathbf{H}_d(\mathbf{m}) \delta \mathbf{d}^{d*}_{l} -  \delta \mathbf{d}^* \|^2_2}{\| \delta \mathbf{d}^* \|^2_2},~~l < l_{max}
\label{eqepsilon1}
\end{equation}
where $l_{max}$ denotes the maximum number of the iteration.
We also introduce 
\begin{equation}
\epsilon_2=\frac{\| \mathbf{H}_d(\mathbf{m}) \delta \mathbf{d}^{d*}_{l} -  \delta \mathbf{d}^* \|^2_2}{\| \mathbf{d}^* \|^2_2},
\label{eqepsilon2}
\end{equation}
to prevent the over-solving of the system when $\delta \mathbf{d}^*$ is small.
To make the proposed algorithm practical, we propose in the next section to estimate a good starting guess of $\delta \mathbf{d}^{d*}_{0}$ (equation~\ref{CG_iter}) from a band-diagonal approximation of the inverse of the data-domain Hessian and mitigate the number of CG iterations accordingly. A more efficient approach to mitigate the number of CG iterations is the design of a good preconditioner of the normal system but this is left for future studies.

\subsection*{Finding approximate data-domain Hessian inverse with matching filters}
\noindent We estimate an approximation of the block matrix $\mathbf{H}_d^{-1}$ \citep[][ Fig.~A-1]{Gholami_2022_EFW} with data-domain matching filters \citep{Liu_2018_ODL}. We remind that the dimension of this Hessian is controlled by the number of time steps and the number of receivers.  Let's first estimate a blurred source extension by time-reverse modeling of the data residuals, that is the adjoint approximation of the least-square problem, equation~\ref{eqwridb}:
\begin{equation}
\delta \mathbf{b}^b=\mathbf{S}(\mathbf{m})^T \mathbf{\delta}\mathbf{d}^{*}.
\label{bd1}
\end{equation}
Solving the forward problem (demigration) with this source extension ($\mathbf{S}(\mathbf{m})\mathbf{\delta}\mathbf{b}^b$) and adding a small damping term to $\mathbf{S}(\mathbf{m})\mathbf{S}(\mathbf{m})^T$ gives the blurred data residuals,
\begin{equation}
\mathbf{\delta}\mathbf{d}^{b} = (\mathbf{S}(\mathbf{m})\mathbf{S}(\mathbf{m})^T + \mu \mathbf{I}) \delta\mathbf{d}^*= \mathbf{H}_d(\mathbf{m}) \delta\mathbf{d}^*. 
\label{bd2}
\end{equation}
From a computational point of view, building the blurring operator requires two simulations (one forward and one backward simulations related to $\mathbf{S}$ and $\mathbf{S}^T$, respectively). Our aim is to find a matching filter $\mathbf{F}$ such that $\mathbf{F} \delta \mathbf{d}^b=\mathbf{F}\mathbf{H}_d(\mathbf{m}) \delta \mathbf{d}^*$ best fits $\delta \mathbf{d}^*$, namely $\mathbf{F} \approx \mathbf{H}_d(\mathbf{m})^{-1}$. We review below different formulations of $\mathbf{F}$.

{\textit{Remark}}: We can determine the value of the penalty parameter $\mu$ from equation~\ref{bd2} as a small fraction of the relative norm of $\mathbf{S}(\mathbf{m})\mathbf{S}(\mathbf{m})^T \delta \mathbf{d}^*$:
\begin{equation}
\mu=\alpha \frac{\| \mathbf{S}(\mathbf{m})\mathbf{S}(\mathbf{m})^T \delta \mathbf{d}^* \|^2_2}{\| \delta \mathbf{d}^* \|_2^2},
\end{equation} 
where $\alpha$ is set to be $0.001-0.01$ in the numerical examples shown later. A better strategy is to define $\mu$ as a small fraction of the maximum eigenvalue of $\mathbf{S(m)S(m)}^T$, while it is difficult to estimate this value with the time-domain formulation. \\

\subsubsection*{SF approximation}

\noindent The SF algorithm relies on a steepest-descent method to solve the scattering-source location problem, equation~\ref{eqwridb}. Accordingly, $\mathbf{F}$ takes the form of a scaled identity matrix, $\mathbf{F}=\gamma \mathbf{I}$, where the step length $\gamma$ is given by
\begin{equation}
\gamma=\frac{ {\delta \mathbf{d}^b}^T \delta \mathbf{d}^* }{ {\delta \mathbf{d}^b}^T \delta \mathbf{d}^b}.
\end{equation}

\subsubsection{1D Wiener matching filter}

\noindent We formulate each receiver-dependent block of $\mathbf{F}$ as a 1D stationary Wiener matching filter (1D-WMF) by minimizing 
\begin{equation}
f_{r}(t)=\argmin_f \frac{1}{2} \int_t \left( {f}_{r}(t) \ast_t  \delta {d}_{{r}}^{b}(t)-{\delta}d_{r}^{*}(t)\right)^2 dt,
\label{eqmf}
\end{equation}
where $\ast_t$ denotes convolution with respect to time; ${f}_{r}(t)$ is the filter that forms the $n_t \times n_t$ Toeplitz block of $\mathbf{F}$ related to receiver $r$; $\delta {d}_{{r}}^{b}(t)$ and ${\delta}d_{r}^*(t)$ are the traces related to receiver $r$ of $\mathbf{\delta}\mathbf{d}^{b}$ and $\mathbf{\delta}\mathbf{d}^*$, respectively. \\
The minimization problem \ref{eqmf} can be solved efficiently in the frequency domain as \citep{Vaseghi_1996_WFF}
\begin{equation}
f_{r} (\omega) =  \frac{ \delta d_{r}^{*} (\omega) \delta \bar{d}_{{r}}^{b} (\omega) }{|\delta d_{{r}}^{b} (\omega)|^2 + \epsilon }, \text{   ~~~~~for all } \omega,
\label{mtf}
\end{equation}
where $\omega$ is the angular frequency, $\bar{\bullet}$ denotes the conjugate of $\bullet$, and $\epsilon > 0$ is a prewhitening parameter. Here, $\mathbf{F}$ is a band diagonal matrix built by arranging $n_r$ Toeplitz matrices along its main diagonal. This 1D WMF approximation lacks the contribution of the off-diagonal blocks of $\mathbf{H}_d^{-1}$, which describes the correlation between receiver traces \citep[][ Fig.~A-1]{Gholami_2022_EFW}. 

\subsubsection*{1D Gabor matching filter}

\noindent Wiener filters are suitable to represent stationary signals, while the seismic traces are non-stationary with respect to time, suggesting that data-domain Hessian cannot be approximated accurately by stationary filters \citep{Yong_2021_IAW}. To alleviate this issue, one can decompose the non-stationary seismic trace into small windows $h(t;\tau)$ to represent the locally-stationary signal:
\begin{equation}
h(t;\tau)= \left( \pi \sigma_t^2 \right)^{-\frac{1}{4}} e^{\frac{-(t-\tau)^2}{2 \sigma_t^2}},
\label{eqmf}
\end{equation}
where $\sigma_t$ controls the length of the time window and $\tau$ is the location of the window center. The solution in the frequency domain is expressed as
\begin{equation}
f_{r} (\omega;\tau) =  \frac{ \delta d_{r}^{*} (\omega;\tau) \delta \bar{d}_{{r}}^{b} (\omega;\tau) }{|\delta d_{{r}}^{b} (\omega;\tau)|^2 + \epsilon }, \text{   ~~~~~for all } \omega ~\text{and} ~ \tau.
\label{mtf}
\end{equation}
The non-stationary filtering of equation \ref{Gabor} can be interpreted as Wiener filtering using the short-time Fourier transform of the whole seismic trace instead of the Fourier transform of selected windows \citep{Gabor_1946_TOC}. Hereafter we refer to this method as the 1D Gabor matching filter (1D-GMF). 

\subsubsection*{2D Gabor matching filter}

\noindent As demonstrated by \citet{Liu_2020_FSL}, 1D trace-by-trace matching filter only approximates the diagonal entries of $\mathbf{H}_d$. To consider the off-diagonal blocks (correlation along the receiver dimension), we devise the following 2D Gabor matching filter (2D-GMF). First, we introduce normalized 2D Gaussian windows \citep{Zhou_2002_BDP}
\begin{equation}
h_{\sigma}(t,r;\tau,l)= \left( \pi \sigma_t \sigma_r \right)^{-\frac{1}{2}} e^{- \left( \frac{(t-\tau)^2}{2 \sigma_t^2} + \frac{(r-l)^2}{2 \sigma_r^2} \right)},
\end{equation}
where $\tau$ and $l$ are the locations of the window center; $\sigma_t$ and $\sigma_r$ control the length of the time window and receiver windows. Their values are determined by trial-and-error method to make the extended data match the observed data. The solution in the frequency-wavenumber domain is given by
\begin{equation} \label{Gabor}
f(\omega,k;\tau,l)= \frac{ \delta d^{*} (\omega,k;\tau,l) \delta \bar{d}^{b} (\omega,k;\tau,l) }{ | \delta d^{b} (\omega,k;\tau,l) |^2 + \epsilon } \text{, ~~~~~for all } (\omega, k)
\end{equation}
where $(\omega,k)$ denotes the 2D frequency-wavenumber space; $\delta d^{*} (\omega,k;\tau,l)$ and $\delta d^{*} (\omega,k;\tau,l)$ are the Fourier domain shot gathers localized at time $\tau$ and trace $\l$, respectively. 

\newpage
\clearpage

\section{Numerical Results}


\subsection*{Marmousi II model}

\subsubsection*{Experimental setup}

\begin{figure}[ht!]
\centering
\includegraphics[width=0.5\linewidth]{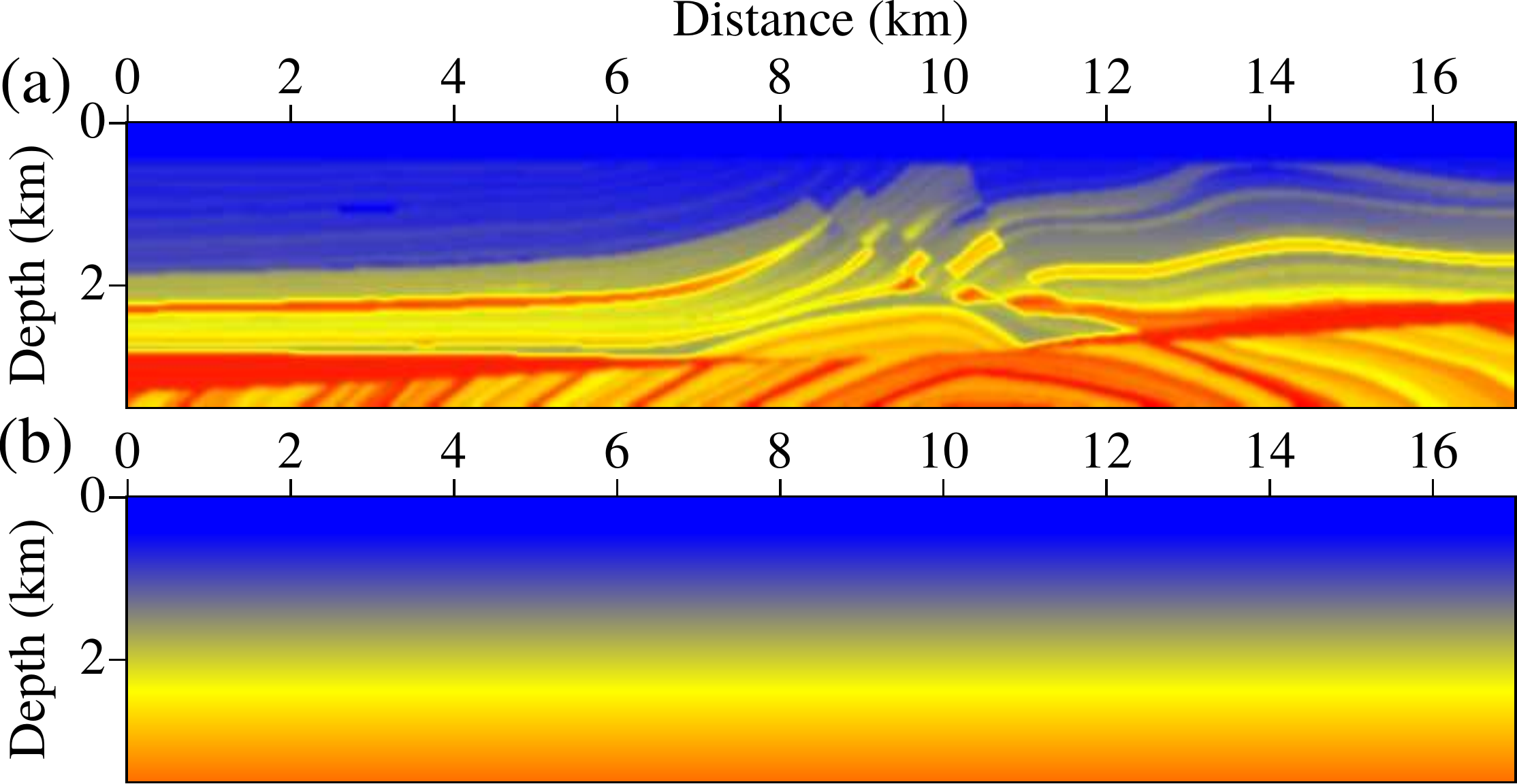}
\caption{(a) Marmousi II model. (b) Initial model.}
\label{Marmousi_true}
\end{figure}

\begin{figure}[ht!]
\centering
\includegraphics[width=1\linewidth]{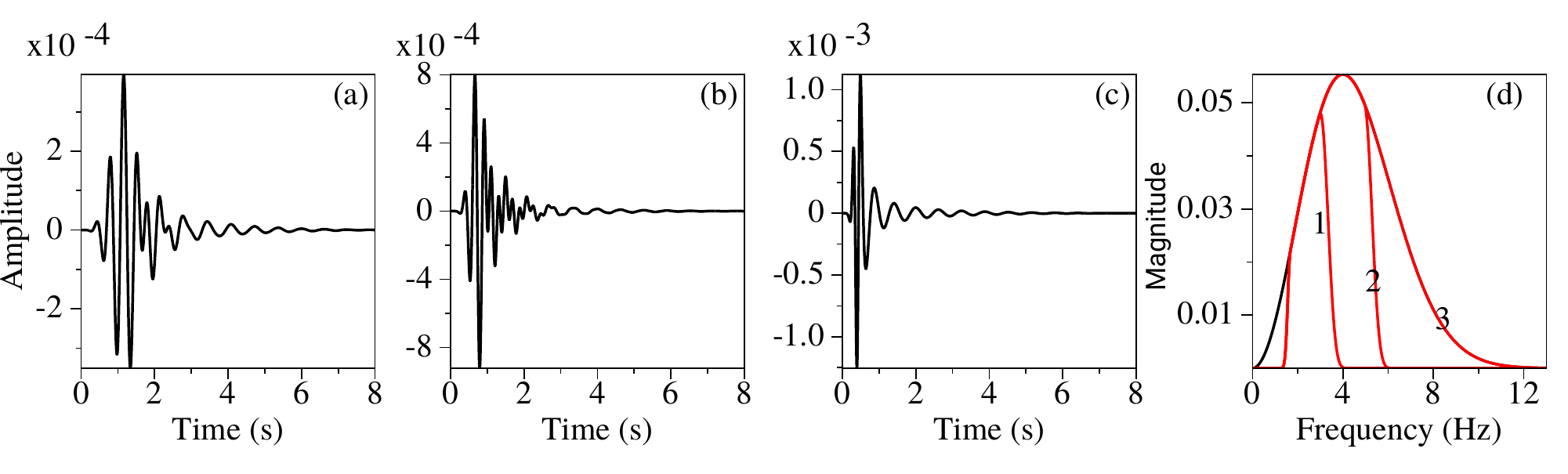}
\caption{Marmousi benchmark. Wavelets for (a) first, (b) second and (c) third multi-scale steps and and their spectral amplitudes (d).}
\label{Marmousi_wavelets}
\end{figure}

\noindent We consider the 17~km $\times$ 3.6~km Marmousi II model (Figure~\ref{Marmousi_true}a). A surface stationary-recording acquisition involves 68 hydrophone receivers on the seabed at a constant depth of 450~m and 227 pressure sources at 50-m depth.  We exploit reciprocity of Green functions to process sources as reciprocal receivers and vice versa to mitigate computational cost. We start the inversion from a 1D gradient-velocity model with velocities ranging from 1.5 to 4~km/s (Figure~\ref{Marmousi_true}b). Wave simulation is performed with a $\mathcal{O}(\Delta t^2, \Delta x^4)$ finite-difference stencil while the sources and the receivers are positioned at arbitrary positions with Kaiser-windowed sinc functions \citep{Hicks_2002_ASR}.  We compute the recorded data (i.e., simulated data in the true model) (Figure~\ref{Marmousi_true}a) with a 14~s recording length and a Ricker wavelet whose peak frequency equals to 4~Hz. Moreover, frequencies smaller than 1.5~Hz are filtered out (Figure~\ref{Marmousi_wavelets}). \\
We perform three series of five tests (Table~\ref{tab1}). We use perfectly matched layer (PML) absorbing condition along the four boundaries of the grid in the first two series, while a free-surface boundary condition is introduced along the top boundary in the third series. Moreover, we regularize the model update with bound constraints using the true values of the minimum and maximum velocities as bounds in the first series, while we apply TV regularization and bound constraints in the last two series. Each test of a series consists of a multiscale-scale inversion  with a frequency continuation strategy using the three band-pass filtered wavelets shown in Figure \ref{Marmousi_wavelets}. At each  multiscale step, we set the spatial grid interval $h$ such that five grid points sample the minimum wavelength, namely $h$=100, 50, and 25~m, respectively.
Each series contains five tests where the data-domain Hessian is accounted for with the SF, 1D-WMF, 1D-GMF, 2D-GMF, and 2D-GMF+CG methods during the first-scale inversion (Table~\ref{tab1}). In the  2D-GMF+CG case, we use the extended data residuals inferred from the 2D-GMF method as a starting guess to perform CG iterations. Then, we build the starting model of the second-scale inversion by up-sampling the final model of the first-scale inversion with cubic-spline interpolation and by smoothing it with a Gaussian filter. The correlation length of the Gaussian smoother is tuned according to the minimum wavelength of the first-scale inversion. We proceed with the second-scale inversion using the same procedure as that used during the first scale. However, we perform IR-WRI with the 2D-GMF method when the first multiscale inversion was performed with the 2D-GMF+CG method to mitigate the computational cost (Table~\ref{tab1}). In the third scale, we perform classical FWI for each test assuming that we are in the linear regime of FWI. 


\begin{table}[htb!]
\begin{center}
\caption{Three-scale inversion tests (Marmousi test). Scale: One scale is associated with a frequency band shown in Figure~\ref{Marmousi_wavelets}. WI: waveform inversion methods (IR-WRI versus FWI). DHM: Data-domain Hessian estimation method (concerns only the first two scales). MSE: Mean-square model error. Three series of tests are performed with/without free surface (FS) boundary condition and with/without TV regularization. The MSE of the final models of each series of tests is provided in the three right-most columns. Series 1: Without FS/without TV (Figures~\ref{Marmousi_inv}-\ref{Marmousi_data_final}); Series 2: Without FS/with TV (Figures~\ref{Marmousi_inv_reg}-\ref{Marmousi_data_final_reg}); Series 3: With FS/with TV (Figures~\ref{Marmousi_inv_ofs}-\ref{Marmousi_data_final_ofs}).}
\small{
\label{tab1}
\begin{tabular}{c|c|c|c|c|c|c|c|}
\cline{2-8} 
& Scale (1/2/3)            &    1      &   2     &  3   & \multicolumn{3}{|c|}{Final model MSE (Marmousi)} \\ \cline{2-8} 
& WI    & \multicolumn{2}{|c|}{IR-WRI} & FWI & noFS/noTV & noFS/TV  & FS/TV \\ \hline 
\multicolumn{1}{|c|}{Test 1} & \multirow{5}{*}{DHM }&    SF     &  SF     & - & 9.32  & 7.49 & 14.11 \\  \cline{1-1} \cline{3-8} 
\multicolumn{1}{|c|}{Test 2} &  &  1D-WMF   & 1D-WMF  & -  &  8.78 & 7.65 & 12.23 \\ \cline{1-1} \cline{3-8} 
\multicolumn{1}{|c|}{Test 3} &  &  1D-GMF   & 1D-GMF  & -  & 8.39 & 6.59  & 9.22 \\ \cline{1-1} \cline{3-8} 
\multicolumn{1}{|c|}{Test 4} &  &  2D-GMF   & 2D-GMF  & -  & 7.98  & 6.80 & 9.20 \\ \cline{1-1} \cline{3-8} 
\multicolumn{1}{|c|}{Test 5} &  &  2D-GMF+CG  &  2D-GMF & - & 7.47 &  6.06 & 7.39 \\ \hline
\end{tabular}
}
\end{center}
\end{table}
%
%
%
%

\subsubsection*{On the impact of the data-domain Hessian on the data fit}

\begin{figure}[ht!]
\centering
\includegraphics[width=1\linewidth]{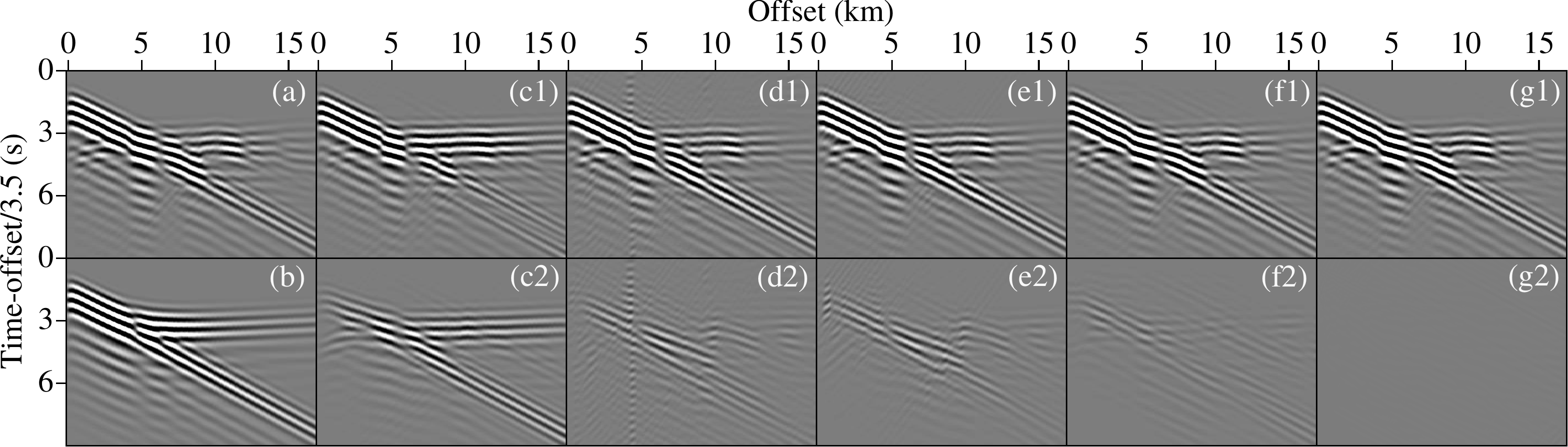}
\caption{(a) Recorded data. (b) Simulated data in initial model (strict solution of wave equation). (c1-g1) Extended-space data computed in initial model using (c1) SF, (d1) 1D-WMF, (e1) 1D-GMF, (f1) 2D-GMF and (g1) CG algorithms. One hundred iterations are performed with the CG method allowing for a quite accurate data fit. (c2-g2) Differences between (a) and (c1-g1). Data are plotted with a reduction velocity (3.5 km/s) and a gain with squared-root of offset.} 
\label{Marmousi_data}
\end{figure}

\begin{figure}[ht!]
\centering
\includegraphics[width=0.7\linewidth]{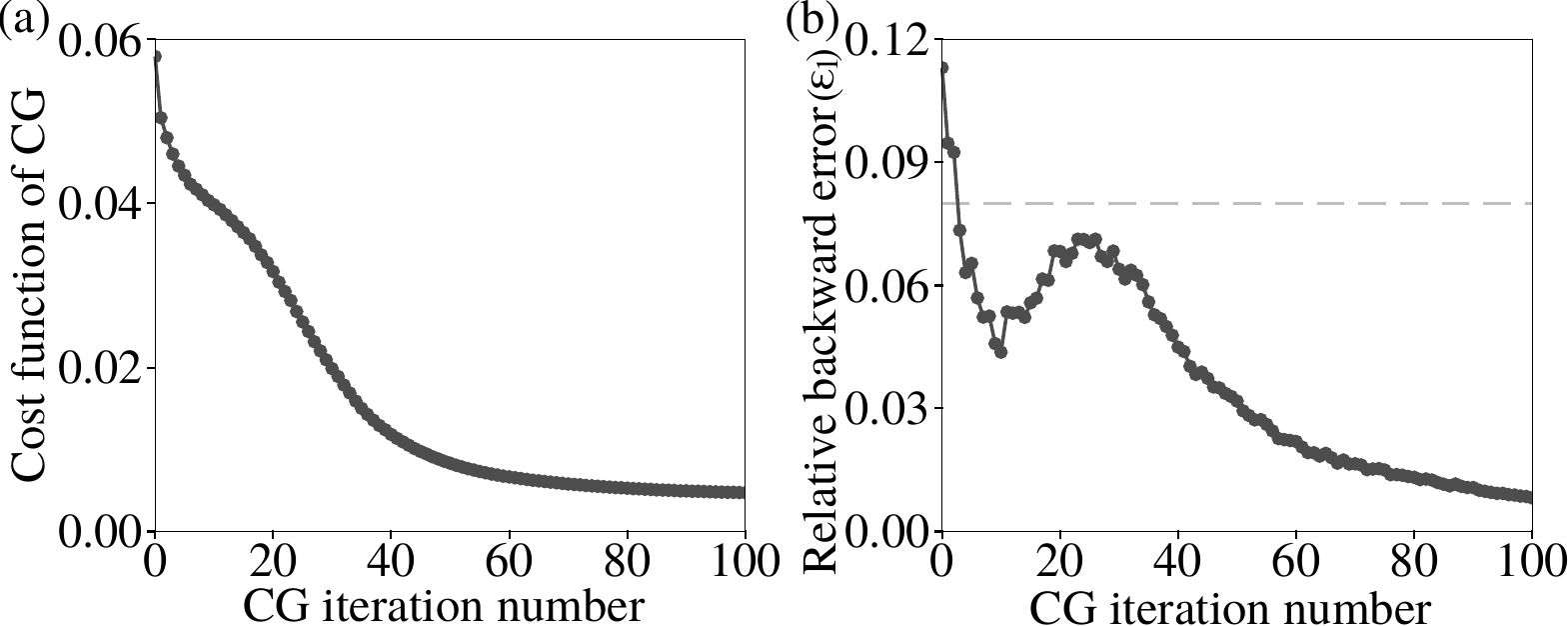}
\caption{Convergence of CG algorithm. (a) Misfit function and (b) relative backward error $\epsilon1$ against number of CG iterations. The dash line shows the relative backward error used for IR-WRI.} 
\label{Marmousi_CG_cost}
\end{figure}

\noindent Figure~\ref{Marmousi_data} illustrates how the accuracy with which the action of $\mathbf{H}_d^{-1}(\mathbf{m})$ on $\delta \bold{d}^*(\bold{m})$ is estimated impacts upon the data fit when a wavefield $\mathbf{u}^e$, equation~\ref{eque3}, is computed in the velocity gradient model shown in Figure~\ref{Marmousi_true}b. In Figure~\ref{Marmousi_data}, the action of $\mathbf{H}_d^{-1}(\mathbf{m})$ on $\delta \bold{d}^*(\bold{m})$ is approximated with the SF, 1D-WMF, 1D-GMF, 2D-GMF, and CG methods. For the CG method, 100 iterations are performed using the result of the 2D-GMF method as a starting guess. The misfit function and the relative backward error $\epsilon_1$ of the CG method against the CG iteration number shows that 100 iterations lead to a relative backward error of the order of 0.01 (Figure~\ref{Marmousi_CG_cost}) and a quite accurate data fit (Figure~\ref{Marmousi_data}g). The inversion tests shown hereafter will show that far less stringent stopping criterion of iteration can be used to mitigate the computational burden of the CG method while still achieving accurate imaging results. As expected, the simulated data with different data-domain Hessian approximations fit the travel times of the recorded data through the migration/demigration of the recorded data \citep{Operto_2022_FBA}. However, the CG method achieves the best amplitude match (Figure~\ref{Marmousi_data}g) while the SF method provides the worst match (Figure~\ref{Marmousi_data}c). Data computed with the 1D-WMF method (Figure~\ref{Marmousi_data}d) better match amplitudes than the SF method but are affected by artifacts. These artifacts are further mitigated by the 1D-GMF and 2D-GMF methods (Figure~\ref{Marmousi_data}(e-f)).

\subsubsection*{On the impact of the data-domain Hessian on extended-source FWI}
\begin{figure}[ht!]
\centering
\includegraphics[width=1\linewidth]{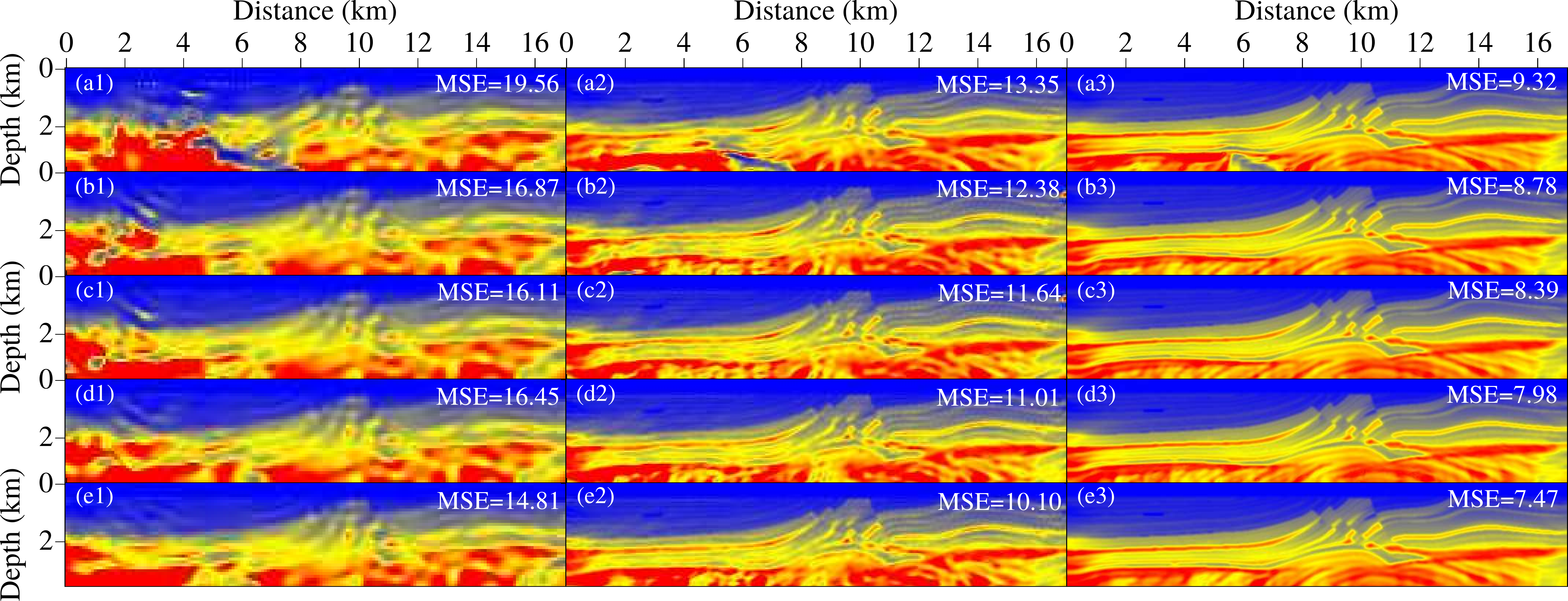}
\caption{IR-WRI/FWI models with different data-domain Hessian approximations. (a) SF, (b) 1D-WMF, (c) 1D-GMF, (d) 2D-GMF and (e) CG+2D-GMF. (a1-e1) First multiscale step. (a2-e2) Second multiscale step. (a3-e3) Third multiscale step. IR-WRI is performed in (a1-e1) and (a2-e2), while FWI is performed in (a3-e3).} 
\label{Marmousi_inv}
\end{figure}

\begin{figure}[ht!]
\centering
\includegraphics[width=0.35\linewidth]{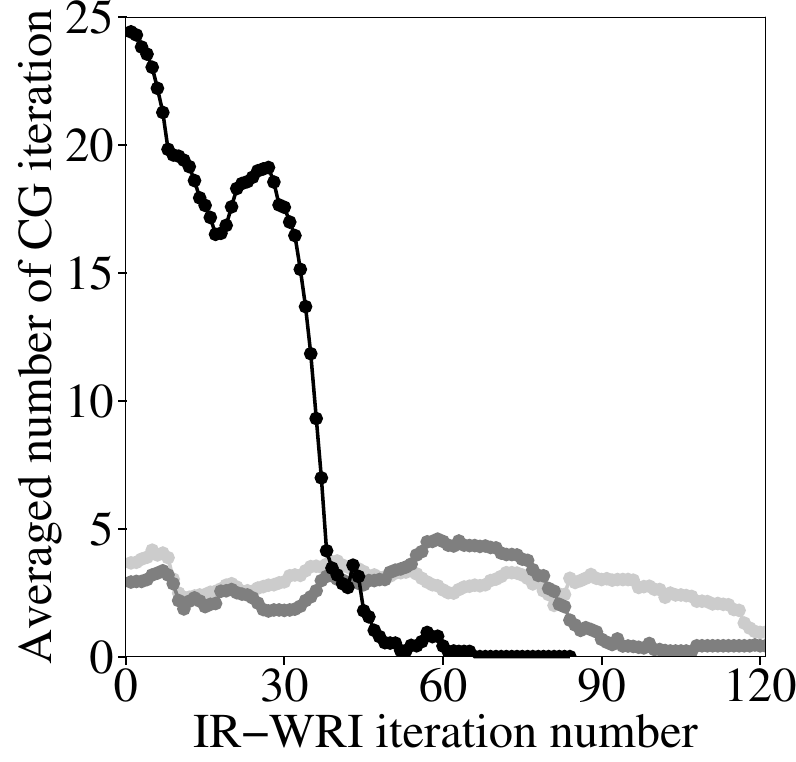}
\caption{Number of CG iterations averaged over sources against the number of IR-WRI iterations (first multiscale step). Light gray: Without free surface, without TV regularization. Dark gray: Without free surface, with TV regularization. Black: With free surface, with TV regularization.} 
\label{Marmousi_CG_num}
\end{figure}

\begin{figure}[ht!]
\centering
\includegraphics[width=0.5\linewidth]{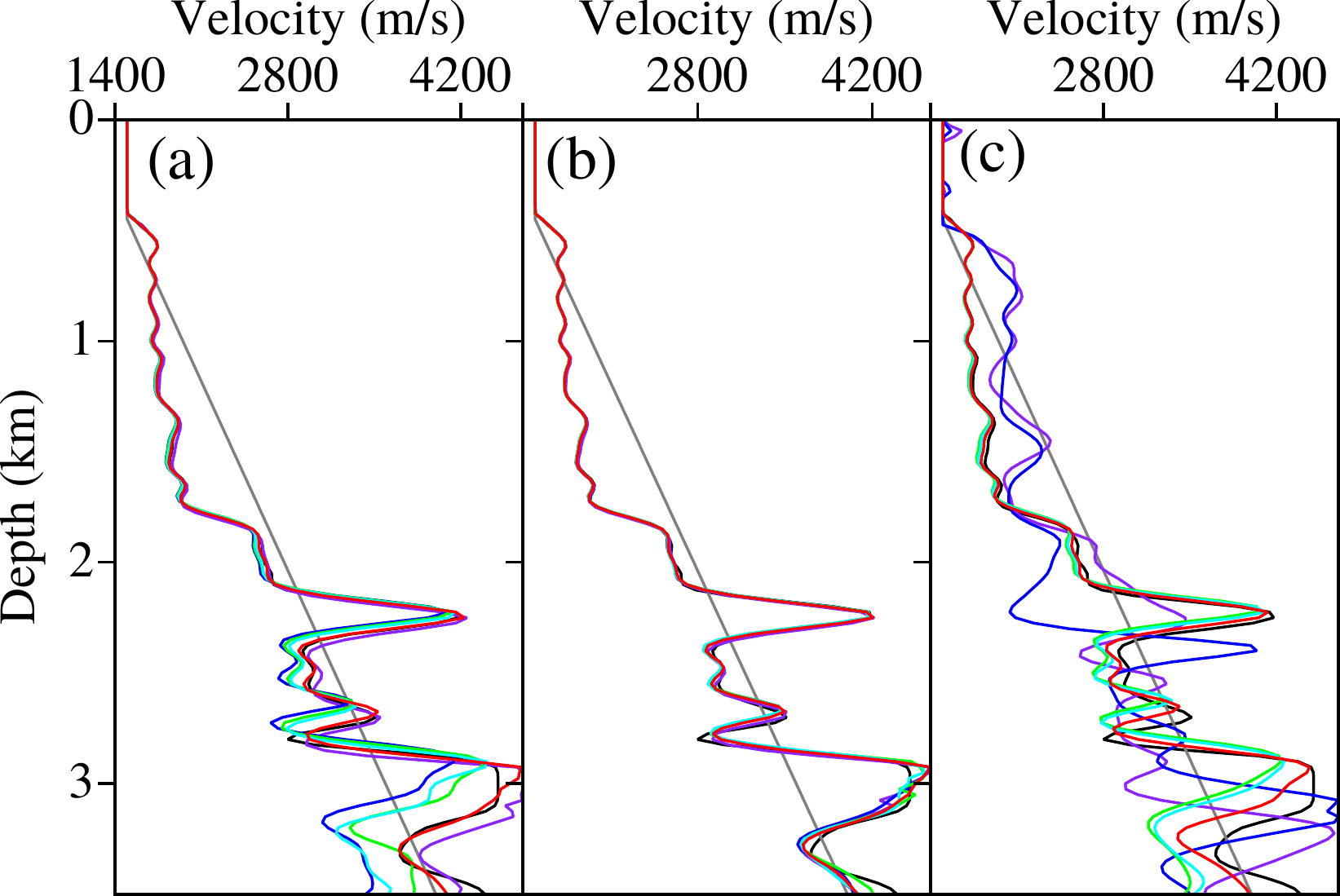}
\caption{Direct comparison between true model (black), initial model (gray) and reconstructed models with SF (purple), 1D-WMF (blue), 1D-GMF (green), 2D-GMF (cyan) and CG+2D-GMF (red) at (a) x=3.75~km. (a) Without free surface, without TV regularization. (b) Without free surface, with TV regularization. (c) With free surface, with TV regularization.} 
\label{Marmousi_logs}
\end{figure}

\begin{figure}[ht!]
\centering
\includegraphics[width=0.7\linewidth]{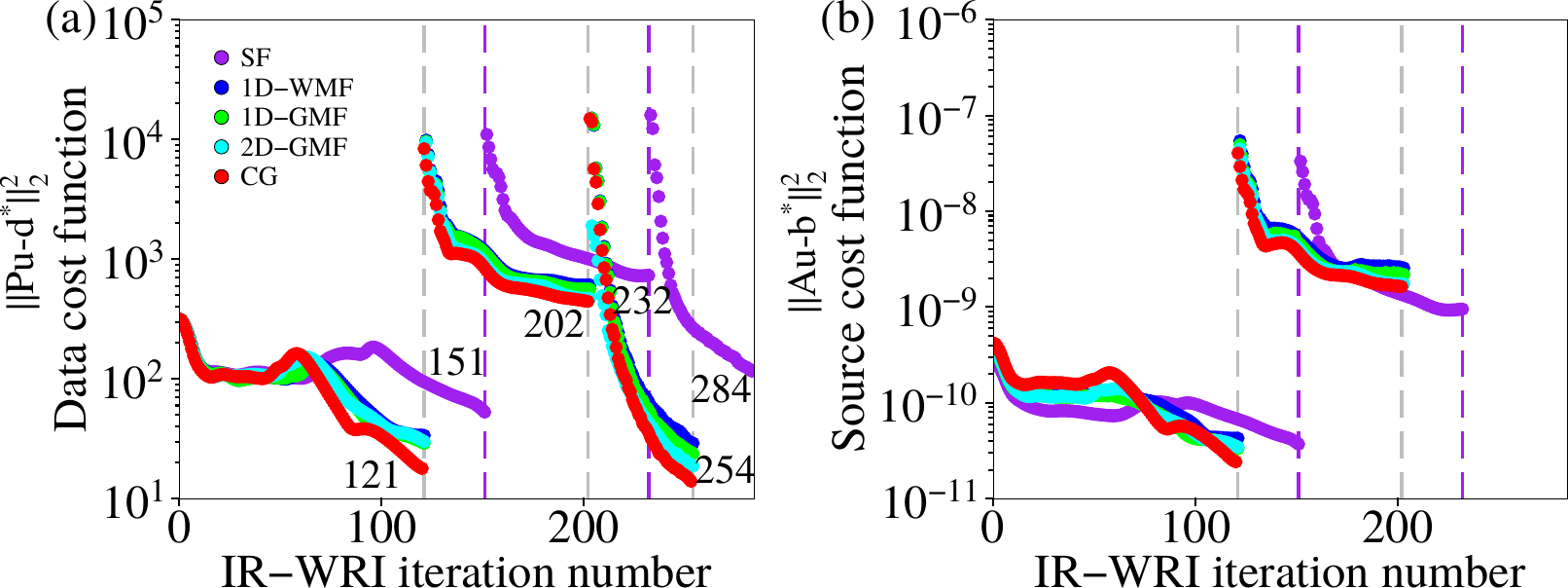}
\caption{(a-b) Data-misfit function  against IR-WRI/FWI iterations. (b) Wave-equation misfit function against IR-WRI iterations. The convergence curves are plotted for the SF (purple), 1D-WMF (blue), 1D-GMF (green), 2D-GMF (cyan) and CG+2D-GMF (red) methods. The dashed lines delineate the three multiscale steps. The number of IR-WRI/FWI iterations performed during each multiscale step are provided. Note that more iterations are performed with the SF method to reach a sufficient decrease of the two misfit functions.} 
\label{Marmousi_cost}
\end{figure}

\begin{figure}[ht!]
\centering
\includegraphics[width=0.8\linewidth]{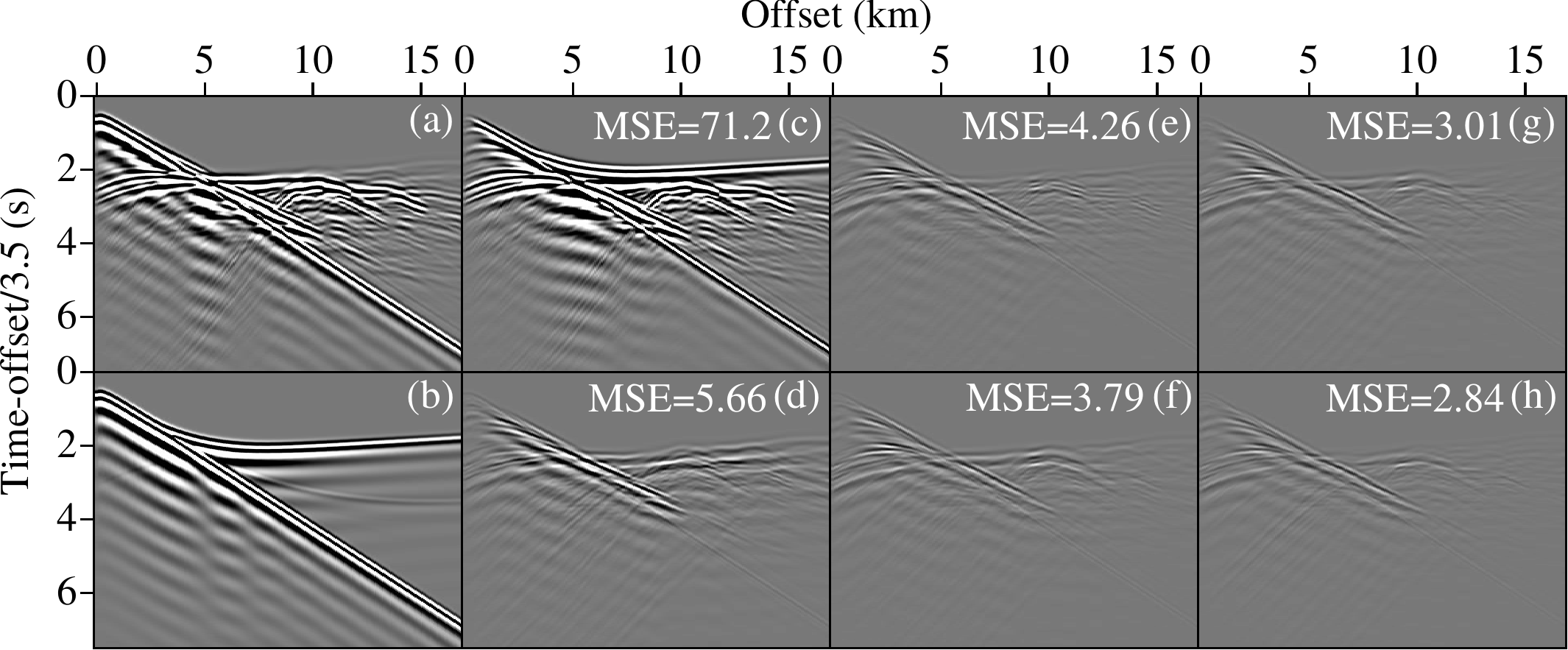}
\caption{Data fit achieved by final FWI models. (a) Recorded data. (b) Simulated data in initial model. (c) Differences between (a) and (b) (Figure~\ref{Marmousi_true}b). (d-h) Differences between (a) and the simulated data in the final FWI models of the third multiscale step (Figure~\ref{Marmousi_inv}(a3-e3)). 
Data are plotted with a reduction velocity (3.5 km/s) and a gain with squared-root of offset.} 
\label{Marmousi_data_final}
\end{figure}

In the first series of tests, we use PML conditions on the surface and we regularize inversion with bound constraints. The reconstructed IR-WRI/FWI models during the three multiscale steps with the different approximations of the data-domain Hessian are shown in Figure~\ref{Marmousi_inv}. 
We use $\epsilon_1=0.08$, equation~\ref{eqepsilon1}, $\epsilon_2=0.02$, equation~\ref{eqepsilon2}, and $l<15$ as a stopping criterion of iteration for the CG method. With this criterion, the number of CG iterations averaged over sources is smaller than five  (Figure~\ref{Marmousi_CG_num}, light gray). Moreover, this number decreases as the inversion approaches the convergence point, which is indeed an important feature to mitigate the computational burden of the method.
The results of the first multiscale step highlight the effect of the data-domain Hessian in the inversion (Figure~\ref{Marmousi_inv}(a1-e1)). The reconstructed model with SF shows artifacts in the left-bottom part of the model (Figure~\ref{Marmousi_inv}(a1)), which were not fully cancelled out after the third FWI multiscale step (Figure~\ref{Marmousi_inv}(a3)). More accurate estimations of the data-domain Hessian with 1D-WMF, 1D-GMF, 2D-GMF, and 2D-GMF+CG continuously mitigate these artifacts during the first multiscale step (Figure~\ref{Marmousi_inv}(b1-e1)). However, the reconstructed models by FWI during the third multiscale step are close in a broad sense for this simple inversion test, no matter 1D-WMF, 1D-GMF, 2D-GMF, or CG method was used in the first step (Figure~\ref{Marmousi_inv}(b3-e3)). Furthermore, a more careful comparison between the true model and the reconstructed ones along a vertical profile at 3.75~km distance shows that the the CG method reconstructs more accurately the high-velocity salt layer at 3~km depth (Figure~\ref{Marmousi_logs})a. The mean-square error (MSE) between the true model and the reconstructed models confirms that the CG method leads to the most accurate solution (Figure~\ref{Marmousi_inv} and Table~\ref{tab1}). \\
The joint course of the data and source misfit functions is illustrated in Figure~\ref{Marmousi_cost}. The data residuals are shown for the three multiscale steps (Figure~\ref{Marmousi_cost}a) while the source residuals are shown for the first two multiscale steps since they are zero during the last FWI step (Figure~\ref{Marmousi_cost}b). The number of IR-WRI/FWI iterations performed during each multiscale step is provided in the figure. Note that the SF method needs more iterations than the other methods to sufficiently decrease the data and the source misfit functions before moving to the last two multiscale steps. Moreover, the CG method generates the most significant source extension during the first iteration of the first multiscale step, which allows for the best initial data fit. The SF method exhibits the opposite trend (Figure~\ref{Marmousi_cost}a, red versus blue curves). At the convergence point of the first multiscale step, the CG method achieves both the lowest data and source misfits (Figure~\ref{Marmousi_cost}a). Finally, the smallest data misfit at the convergence point of the final FWI step is reached when the CG method was used during the first IR-WRI step (Figure~\ref{Marmousi_cost}c, red curve). This improved data fit provided by the CG iterations is further illustrated in Figure~\ref{Marmousi_data_final}, which shows the recorded data and the differences with the simulated data in the final model of the last FWI step for the SF, 1D-WMF, 1D-GMF, and 2D-GMF, and CG algorithms.


%
%
%
%
%
%
%
%

\newpage
\clearpage

\subsubsection*{On the impact of TV regularization on extended-source FWI}

\begin{figure}[ht!]
\centering
\includegraphics[width=1\linewidth]{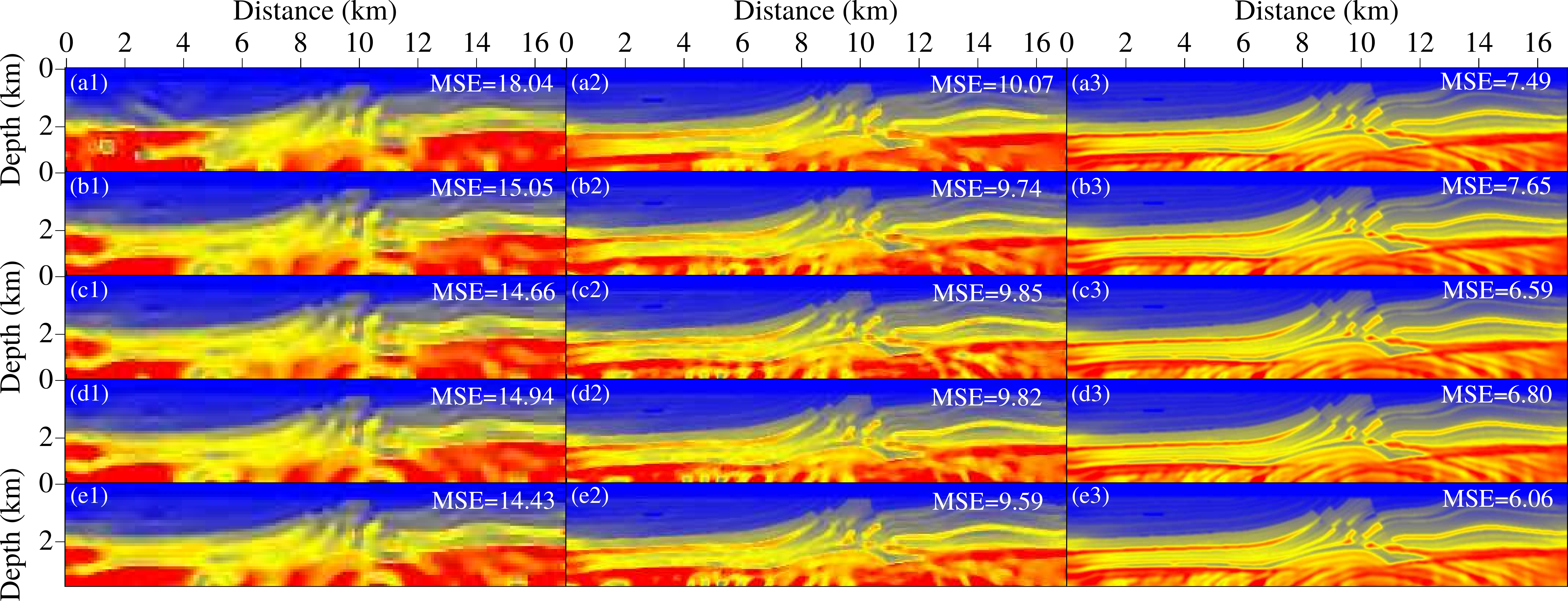}
\caption{IR-WRI/FWI models. Same as Figure~\ref{Marmousi_inv} when TV regularization is applied during the first two multiscale steps.} 
\label{Marmousi_inv_reg}
\end{figure}

\begin{figure}[ht!]
\centering
\includegraphics[width=0.8\linewidth]{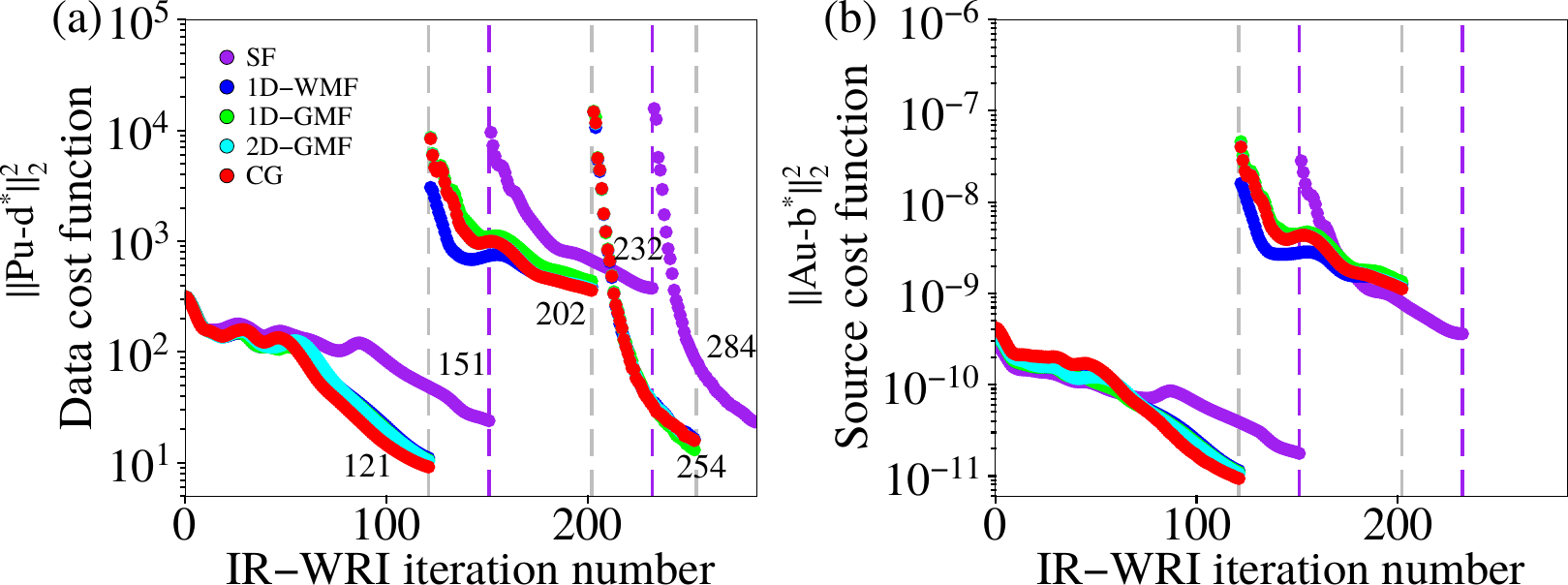}
\caption{Cost functions. Same as Figure~\ref{Marmousi_cost} when TV regularization is applied during the first two multiscale steps.} 
\label{Marmousi_cost_reg}
\end{figure}

\begin{figure}[ht!]
\centering
\includegraphics[width=0.8\linewidth]{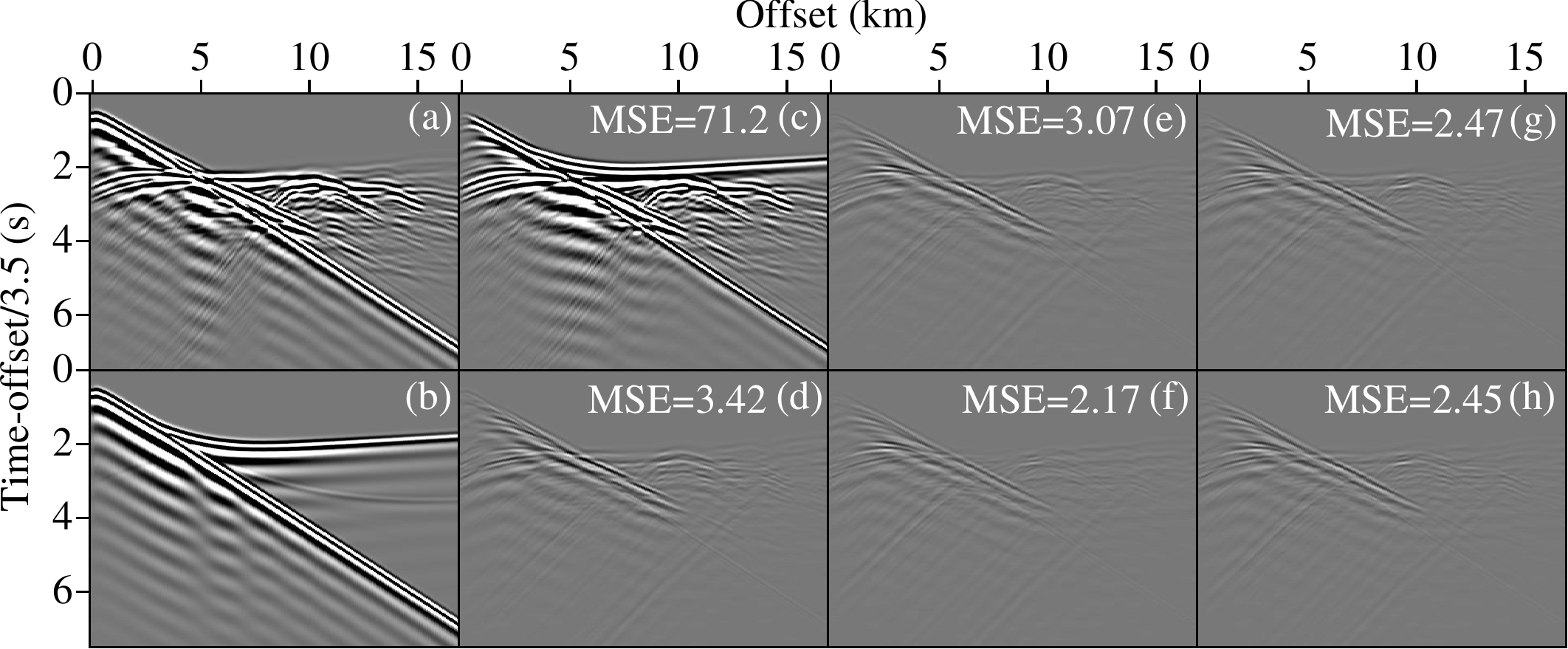}
\caption{Data fit achieved by final FWI models. Same as Figure~\ref{Marmousi_data_final} when TV regularization is applied during the first two multiscale steps.} 
\label{Marmousi_data_final_reg}
\end{figure}

\noindent In the second series of tests, we apply sparsity-promoting TV regularization during model estimation, still without considering free-surface multiples. We use the same experimental setup as the one of the previous section. We apply the TV regularization during the first and second multiscale steps, while FWI is performed without regularization during the last multiscale step to foster data fit. The reconstructed models during the three multiscale steps (Figure~\ref{Marmousi_inv_reg}) can be compared with those obtained without regularization  (Figure~\ref{Marmousi_inv}). The number of CG iterations against the number of IR-WRI iterations  during the first multiscale step shows that the regularization may require to perform more CG iterations compared to the regularization-free algorithm (Figure~\ref{Marmousi_CG_num}, light versus dark gray curves). However, the maximum number of CG iterations remains lower than 5.
Compared with the results obtained without regularization (Figure~\ref{Marmousi_inv}), TV regularization effectively removes noise and artifacts generated by the salt layer in the left part of the model. 
This comment also applies to the SF method, from which an accurate final model is built by FWI during the last multiscale step. This statement is further validated by the direct comparison between the true model and the reconstructed ones along the vertical profile of Figure~\ref{Marmousi_logs}b, which shows a similar accuracy of the final FWI models. However, a more quantitative assessment provided by the MSE shows that the the CG method still leads to the most accurate model (Figure~\ref{Marmousi_inv_reg} and Table~\ref{tab1}). \\
The convergence curves of the data and source misfit functions show that the SF method still generates the less energetic source extension at the first iteration (Figure~\ref{Marmousi_cost_reg}). However, it reaches a data fit at the convergence point of the first multiscale step that is close to that of the other methods. We note that the TV regularization allows for a more efficient decrease of the source misfit during the first two multiscale steps (compare Figures~\ref{Marmousi_cost}a and ~\ref{Marmousi_cost_reg}a). The similar convergence curves of the last multiscale step further supports that all of the methods generate final FWI models of similar accuracy (Figure~\ref{Marmousi_cost_reg}c).
For sake of completness, Figure~\ref{Marmousi_data_final_reg} shows the differences between the recorded data and the simulated data in the final FWI models. The MSEs confirms that the final data fit achieved by the different algorithms (from 3.42 to 2.45) is more homogeneous compared to the regularization-free application (from 5.66 to 2.84) (Figure~\ref{Marmousi_data_final}). Moreover, the final data fit achieved with the best algorithm has been improved when regularization is used (2.17 versus 2.84). Note that the 1D-GMF method provides slightly more accurate datafit than the CG method, which further highlights that the regularization tends to level the contribution of the data-domain Hessian estimation  methods for this relatively simple benchmark.


\newpage
\clearpage

\subsubsection*{On the impact of the free surface on extended-source FWI}

\begin{figure}[ht!]
\centering
\includegraphics[width=1\linewidth]{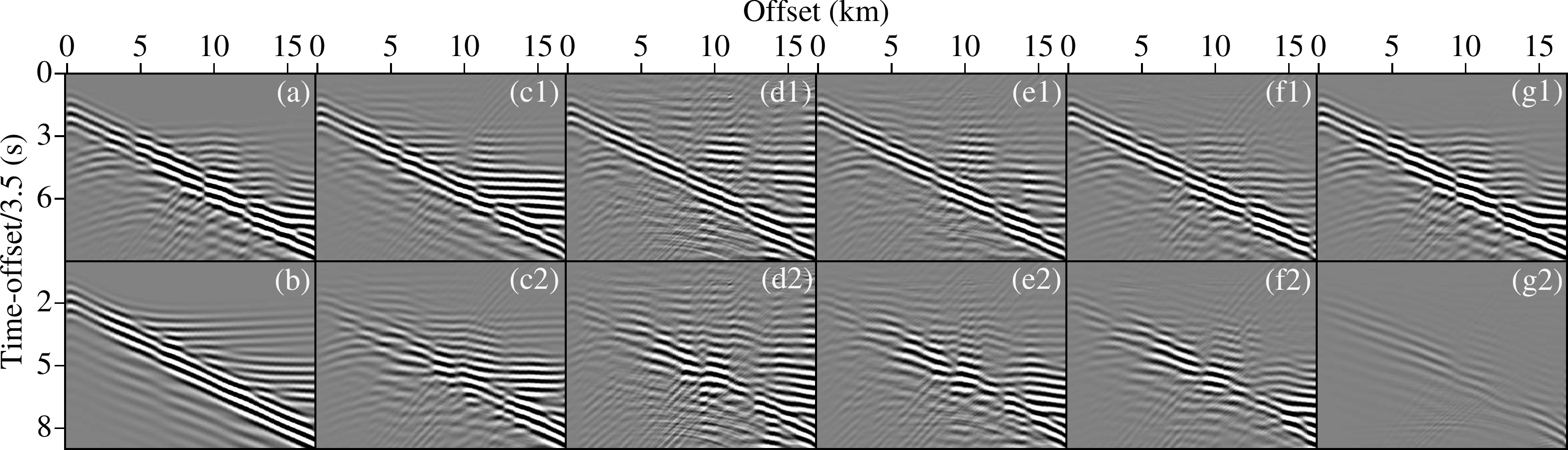}
\caption{Marmousi test with free surface. (a) True data. (b) Reduced-space data. (c1-g1) Extended data using (c1) SF, (d1) 1D-WMF, (e1) 1D-GMF, (f1) 2D-GMF and (g1) CG with 100 iterations. (c2-g2) Differences between (a) and (c1-g1). Data are plotted with a reduction velocity (3.5 km/s) and a gain with squared-root of offset.} 
\label{Marmousi_data_ofs}
\end{figure}

\begin{figure}[ht!]
\centering
\includegraphics[width=0.6\linewidth]{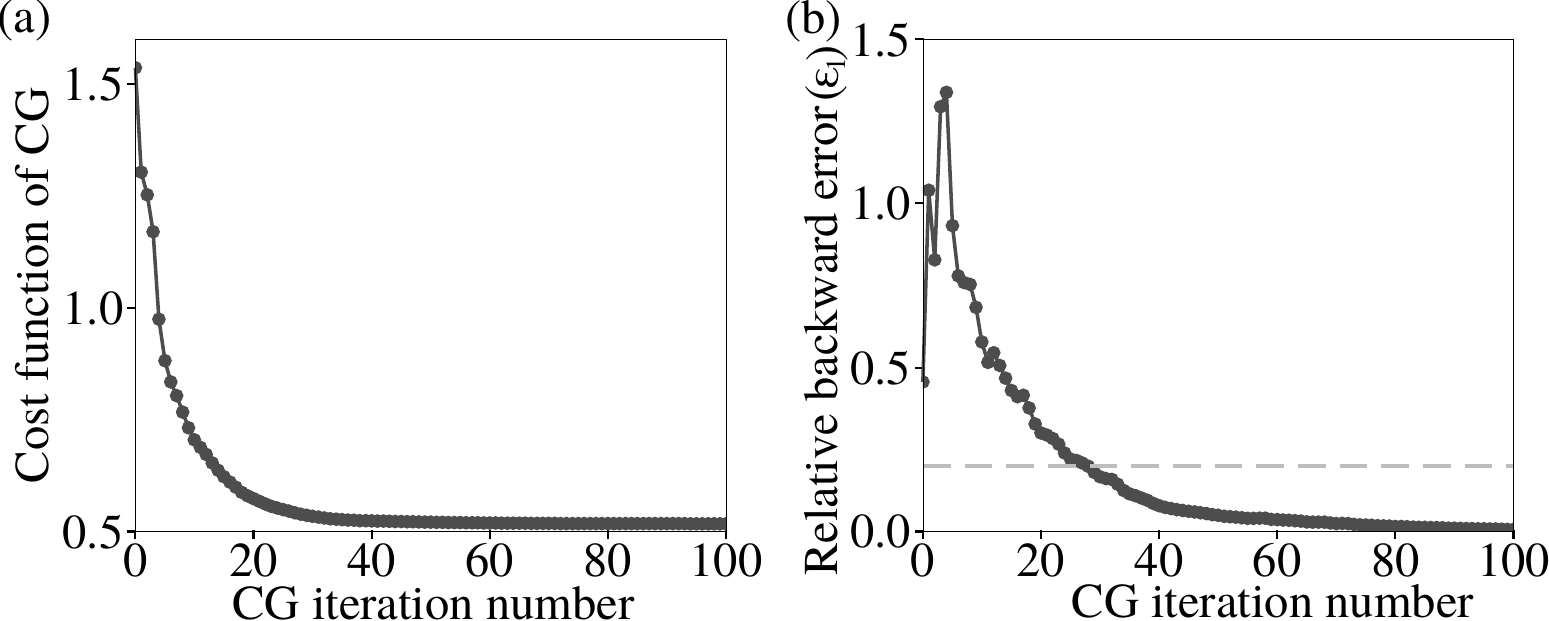}
\caption{Convergence of CG algorithm. (a) Misfit function and (b) relative backward error $\epsilon1$ against number of CG iterations. The dash line shows the relative backward error used for IR-WRI.} 
\label{Marmousi_CG_cost_ofs}
\end{figure}

\begin{figure}[ht!]
\centering
\includegraphics[width=1\linewidth]{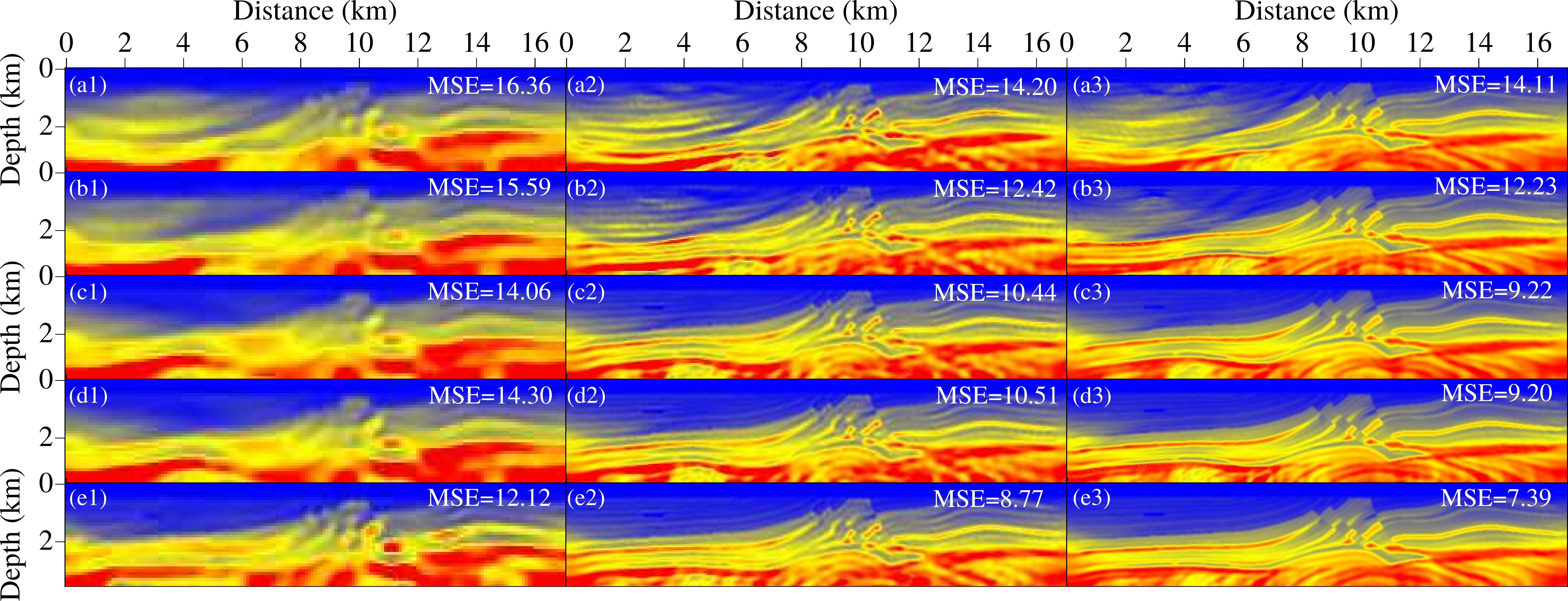}
\caption{IR-WRI/FWI models. Same as Figures~\ref{Marmousi_inv} and \ref{Marmousi_inv_reg} when free surface boundary condition is used on top of the model.} 
\label{Marmousi_inv_ofs}
\end{figure}

\begin{figure}[ht!]
\centering
\includegraphics[width=0.8\linewidth]{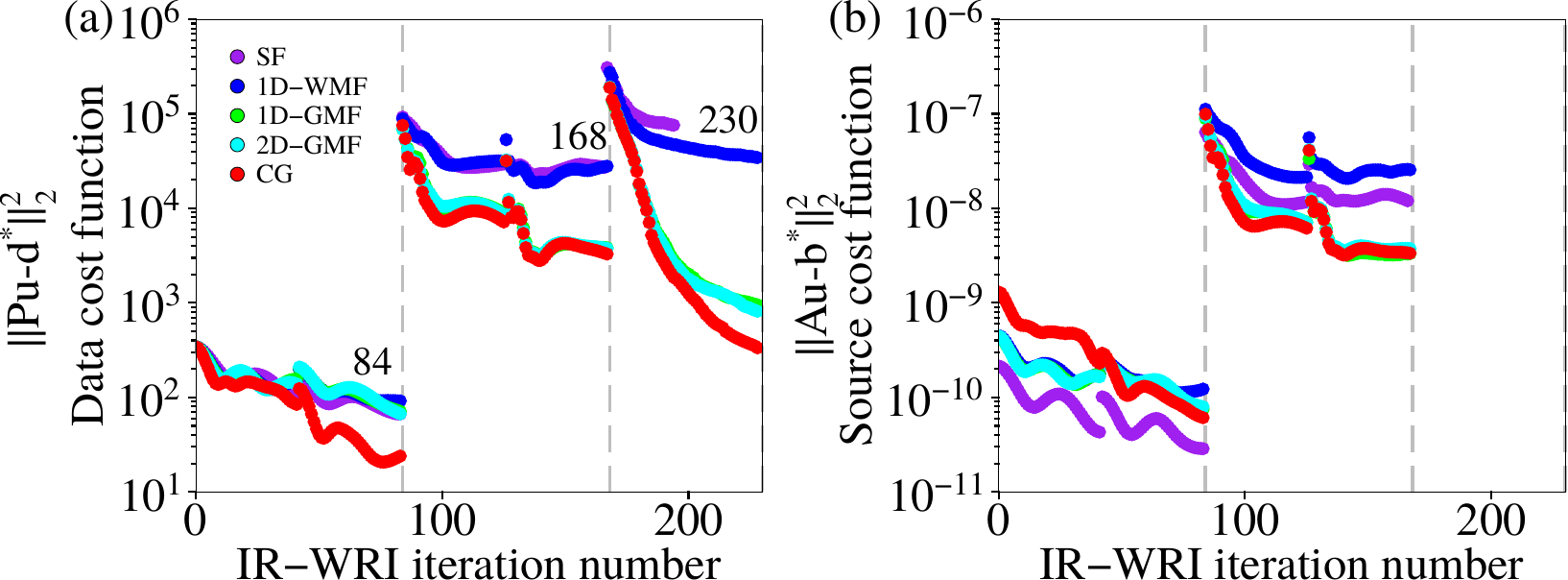}
\caption{Cost functions. Same as Figures~\ref{Marmousi_cost} and~\ref{Marmousi_cost_reg} when free surface boundary condition is used on top of the model.} 
\label{Marmousi_cost_ofs}
\end{figure}

\begin{figure}[ht!]
\centering
\includegraphics[width=0.8\linewidth]{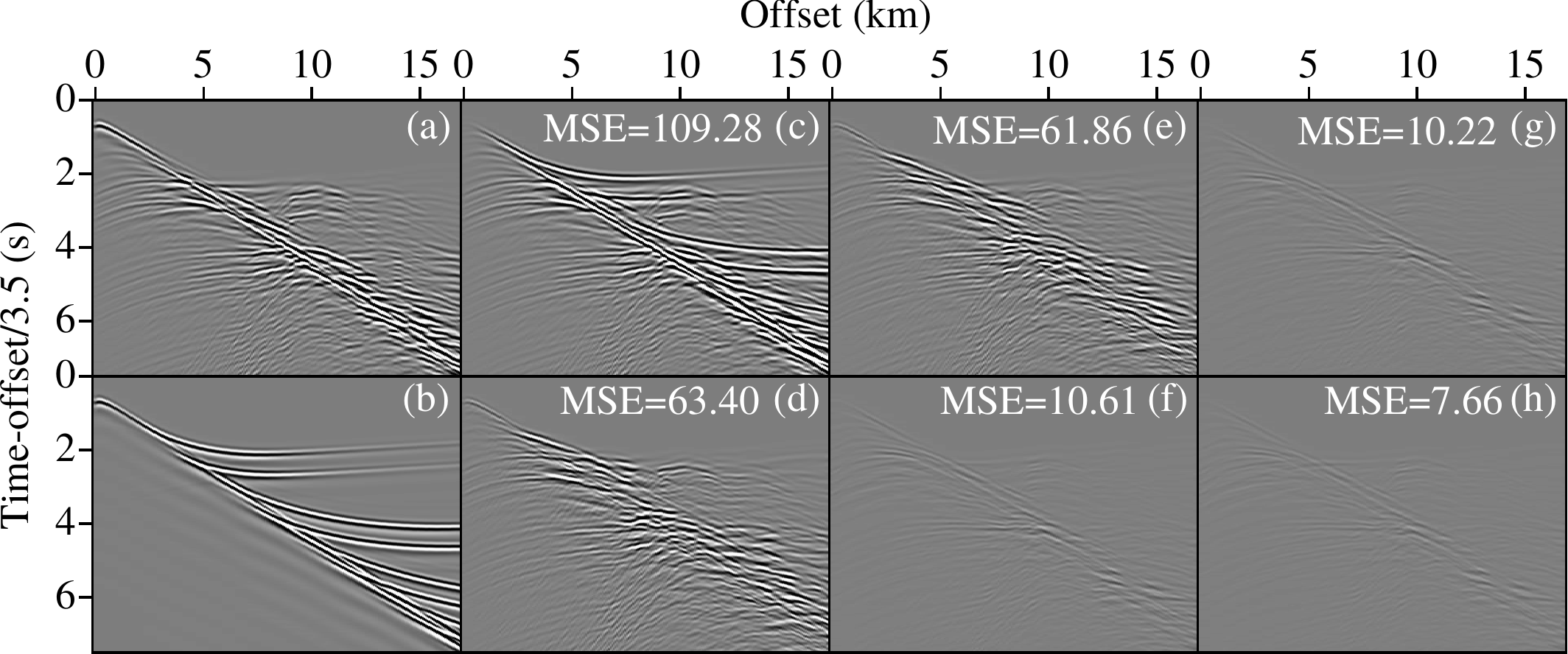}
\caption{Data fit achieved by final FWI models. Same as Figures~\ref{Marmousi_data_final} and \ref{Marmousi_data_final_reg}  when free surface boundary condition is used on top of the model.} 
\label{Marmousi_data_final_ofs}
\end{figure}

\noindent The third series of tests tackles a more challenging problem where the free-surface boundary condition is implemented on the surface of the Marmousi II model. We remind that we apply TV regularization during this series of tests (Table~\ref{tab1}). We first check the effect of the free surface on the data fitting by wavefield reconstruction. Comparing the recorded data with and without free-surface boundary condition (Figures~\ref{Marmousi_data_ofs}a and Figure~\ref{Marmousi_data}a, respectively) highlights the wavefield complexity introduced by the free surface, in particular at long offsets, which makes the data matching more challenging. Figure~\ref{Marmousi_data_ofs} shows how the extended data gradually match better the recorded counterpart as the data-domain Hessian is taken into account more accurately. Figure~\ref{Marmousi_CG_cost_ofs} shows that the more complex data anatomy generated by the free-surface condition dramatically slows the convergence of the CG method (compare Figures~\ref{Marmousi_CG_cost} and \ref{Marmousi_CG_cost_ofs}). \\
The reconstructed models during the three multiscale steps with the different data-domain Hessian estimation methods are shown in Figure~\ref{Marmousi_inv_ofs}. 
We use $\epsilon_1=0.2$, $\epsilon_2=0.2$ and $l<15$ as stopping criterion of iteration for the CG method. These values are significantly higher than those used in the previous test to keep the computational overhead generated by CG iterations manageable. This indeed implies that the effects of the data-domain Hessian are not taken into account as accurately as in the previous test. \\
A comparison with the previous results obtained with absorbing boundary conditions (Figure~\ref{Marmousi_inv_reg}) illustrates the substantial increase of non-linearity and ill-posedness introduced by the free-surface reflections in IR-WRI. Artifacts take the form of a shallow high-velocity patch in the left part of the model, which generates mispositioning of the salt layer in the deeper part. Consistently with the initial data fit illustrated in Figure~\ref{Marmousi_data_ofs}, these artifacts are gradually mitigated  as the data-domain Hessian is taken into account more accurately during the first two multiscale steps. Acceptable results are obtained with the CG method although a small footprint of this artifact remains in the final FWI model (Figure~\ref{Marmousi_data_ofs}(e3)). The number of CG iteration is gradually decreased during the first-scale inversion (Figure~\ref{Marmousi_CG_num}, black curve). However, up to 25 iterations are necessary during the first iterations of the first multiscale step. This sensitivity of IR-WRI to the accuracy of the data-domain Hessian estimation is further illustrated by the vertical log shown in Figure~\ref{Marmousi_logs}c. The paths followed by the data and source misfit functions against iterations are shown in Figure~\ref{Marmousi_cost_ofs} and can be compared with those of Figure~\ref{Marmousi_cost_reg}. As in the previous tests, the source extensions estimated by SF are greatly underestimated during the early iterations compared to those estimated by the matching filter and CG methods. This underestimation traps SF into a  local minimum when free-surface multiples complicate the data. The quite different convergence behavior  of FWI during the third multiscale step further highlights the significant heritage of the data-domain Hessian estimation during the first two multiscale steps (Figure~\ref{Marmousi_cost_ofs}c which can be compared with Figure~\ref{Marmousi_cost_reg}c).  The simulated gathers in the final FWI models can be compared with the recorded counterpart in Figure~\ref{Marmousi_data_final_ofs}. Data fit achieved when 1D-GMF, 2D-GMF and CG methods are acceptable although the data residuals are higher than in Figure~\ref{Marmousi_data_final_reg}, while the data fit achieved by the SF and 1D-WMF methods are not acceptable.
%
%
%
%
%
%
%
%
%
%

\newpage
\clearpage


\subsection*{BP salt model}

\subsubsection*{Experimental setup}

\begin{figure}[ht!]
\centering
\includegraphics[width=0.6\linewidth]{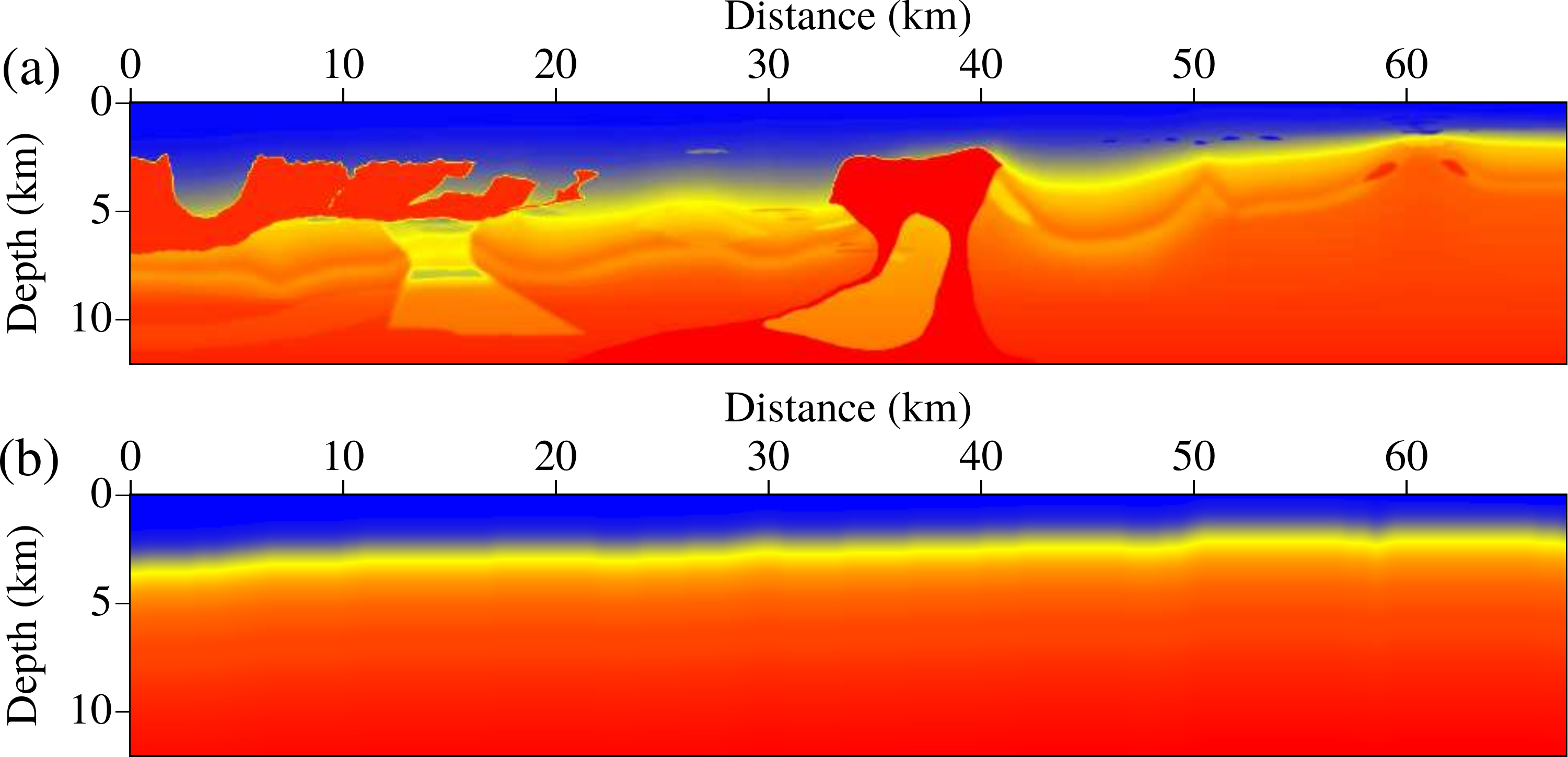}
\caption{(a) BP salt model. (b) Initial model.} 
\label{BP_true_init}
\end{figure}

\begin{figure}[ht!]
\centering
\includegraphics[width=1\linewidth]{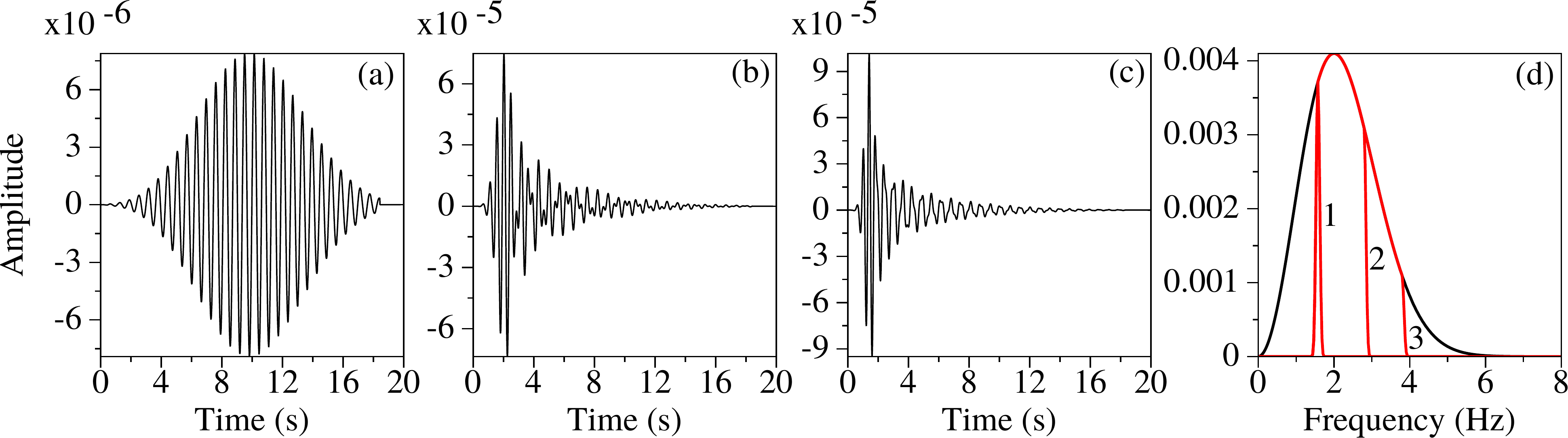}
\caption{BP salt benchmark. Wavelets for (a) first, (b) second and (c) third multi-scale steps and and their spectral amplitudes (d).} 
\label{BP_wavelets}
\end{figure}

\noindent We now consider the full 2004 BP-salt model \citep{Billette_2004_BPB} to assess our method in presence of large velocity contrasts and complex salt bodies (Figure~\ref{BP_true_init}a). We design a surface stationary recording acquisition with 168 hydrophones evenly deployed along the bathymetry and 671 pressure sources at 100-m depth. In this group test, PMLs are applied along the four boundaries of the model. Free-surface multiples are difficult to manage on this challenging model and may require the implementation of layering-stripping strategies in the inversion or more accurate data-domain Hessian estimation as discussed in the next section. We compute the recorded data in the true model (Figure~\ref{BP_true_init}a) with a recording time of 27~s and a Ricker wavelet whose peak frequency equals to 2~Hz. Frequencies below 1.5~Hz are filtered out (Figure \ref{BP_wavelets}). The starting model is a crude velocity gradient model shown in Figure~\ref{BP_true_init}b.  We perform a multi-scale inversion with a frequency continuation strategy using the three wavelets shown in Figure \ref{BP_wavelets}. The spatial grid intervals during each multiscale steps are  150, 100, and 75~m, respectively. In a similar way to the Marmousi setup outlined in Table~\ref{tab1}, we perform IR-WRI with the SF, 1D-WMF, 1D-GMF, 2D-GMF and CG + 2D-GMF methods during the first and second multiscale steps, while we perform classical FWI during the third step. We perform two series of tests with and without TV regularization, while the tests involving free-surface multiples are left for future studies.
%
%
%
%

\subsubsection*{On the impact of the data-domain Hessian on the data fit}

\begin{figure}[ht!]
\centering
\includegraphics[width=1\linewidth]{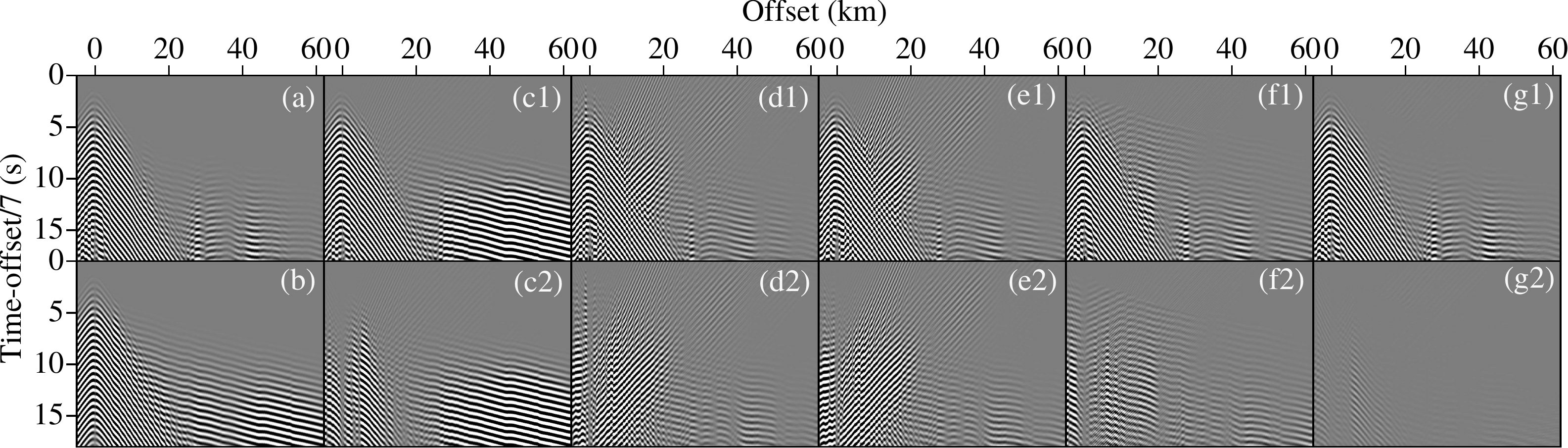}
\caption{Data fit in the initial model. (a-b) Reduced-space data computed in (a) true model and (b) initial model. (c1-g1) Extended-space data computed in initial model with (c1) SF, (d1) 1D-WMF, (e1) 1D-GMF, (f1) 2D-GMF and (g1) CG methods. One hundred CG iterations are performed. (c2-g2) Differences between (a) and (c1-g1). Data are plotted with a reduction velocity (7 km/s) and a gain with squared-root of offset.} 
\label{BP_data}
\end{figure}

\begin{figure}[ht!]
\centering
\includegraphics[width=0.7\linewidth]{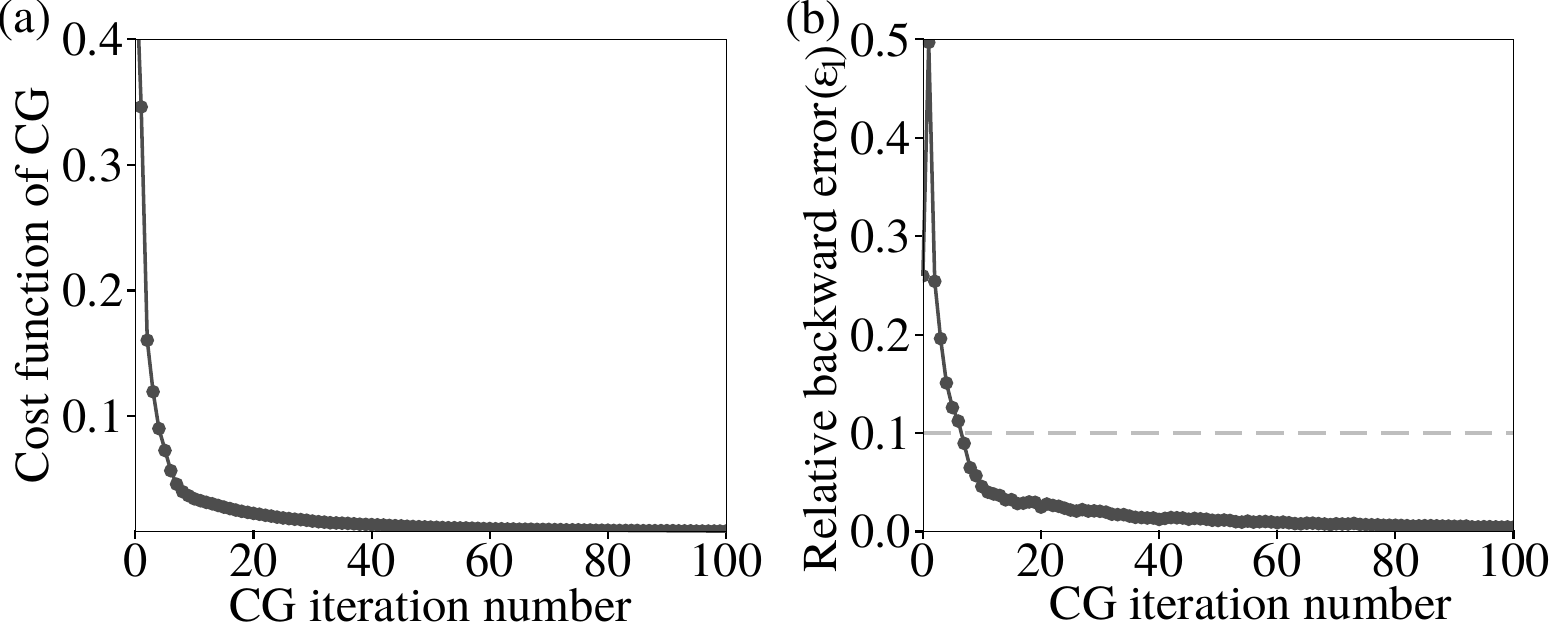}
\caption{Convergence of CG algorithm. (a) Cost function and (b) relative backward error against CG iteration number. Dashed line shows relative backward error used for IR-WRI.} 
\label{BP_CG_cost}
\end{figure}

\noindent As in previous tests, we first check the impact of the accuracy of the data-domain Hessian on the data fit when $\bold{u}^e$ is computed in the starting model (Figure~\ref{BP_true_init}b). Figure~\ref{BP_data}(a-b) shows obvious cycle skipping between the recorded data and the simulated data in the initial model when the wave equation is strictly satisfied. The simulated data in the extended space are shown in Figure~\ref{BP_data}(c-e). For the CG method, 100 iterations are performed using the result of the 2D-GMF method as a starting guess. The SF algorithm generates simulated data that don't fit the recorded data well, especially at long offsets (Figure~\ref{BP_data}c). These mismatches are greatly reduced with 1D matching filter methods (1D-WMF and 1D-GMF) (Figure~\ref{BP_data}(d-e)). However, these methods also introduce coherent linear artifacts. More accurate results are obtained with 2D-GMF (Figure~\ref{BP_data}f), which are further improved with 100 CG iterations (Figure~\ref{BP_data}g). The CG misfit function and the relative backward error against the 100 CG iterations highlight the convergence of the CG algorithm toward accurate solution (Figure~\ref{BP_CG_cost}) as illustrated by the small data residuals in Figure~\ref{BP_data}g. 
\subsubsection*{The impact of the data-domain Hessian on extended-source FWI}

\begin{figure}[ht!]
\centering
\includegraphics[width=0.6\linewidth]{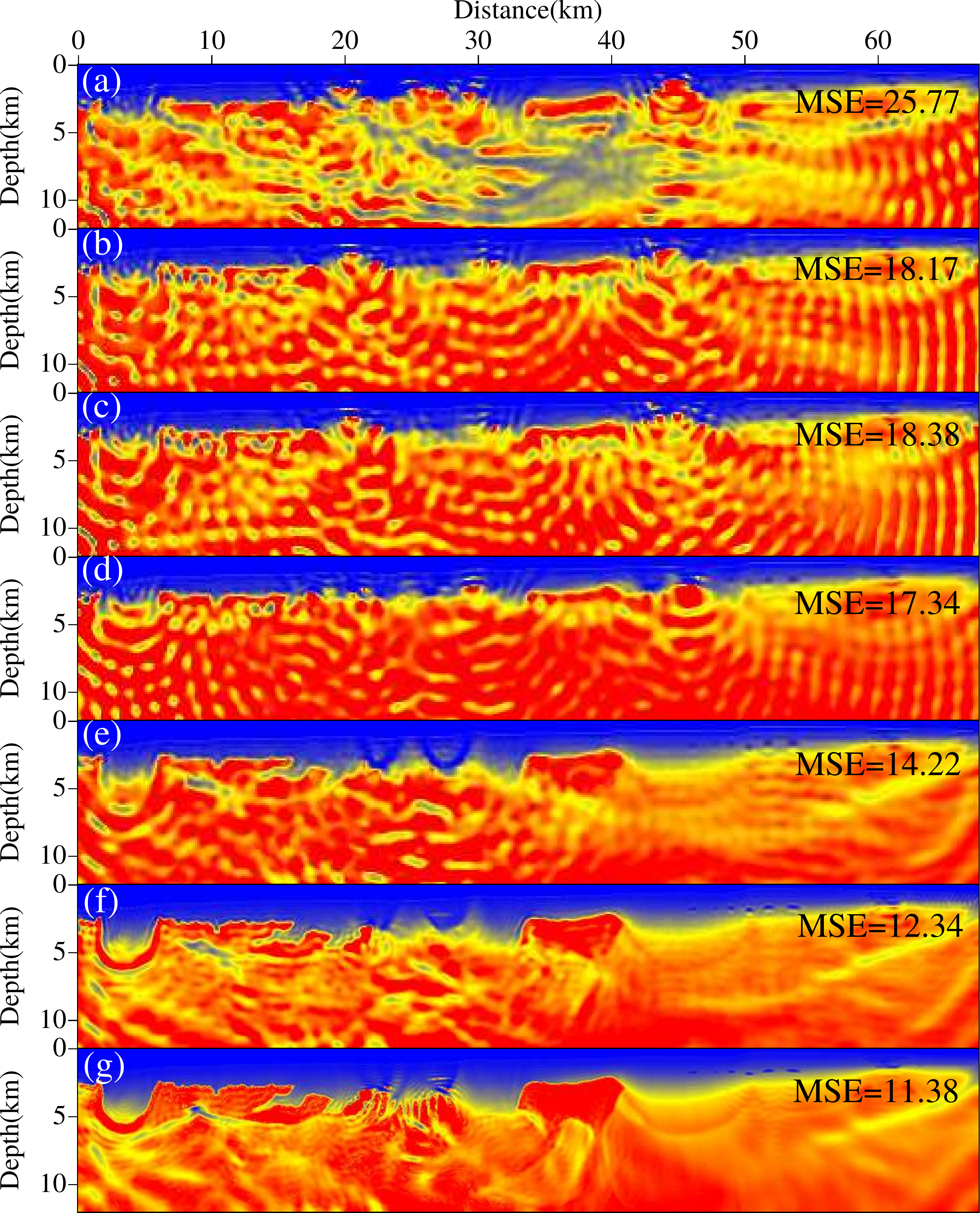}
\caption{(a-e) IR-WRI results close of the first multiscale step using (a) SD, (b) 1D-WMF, (c) 1D-GMF, (d) 2D-GMF and (e) CG + 2D-GMF methods. (f) IR-WRI model close of the second multiscale step starting from the model shown in (e). (g) FWI model after the third multiscale step starting from the velocity model shown in (f).} 
\label{BP_inv}
\end{figure}

\begin{figure}[ht!]
\centering
\includegraphics[width=0.4\linewidth]{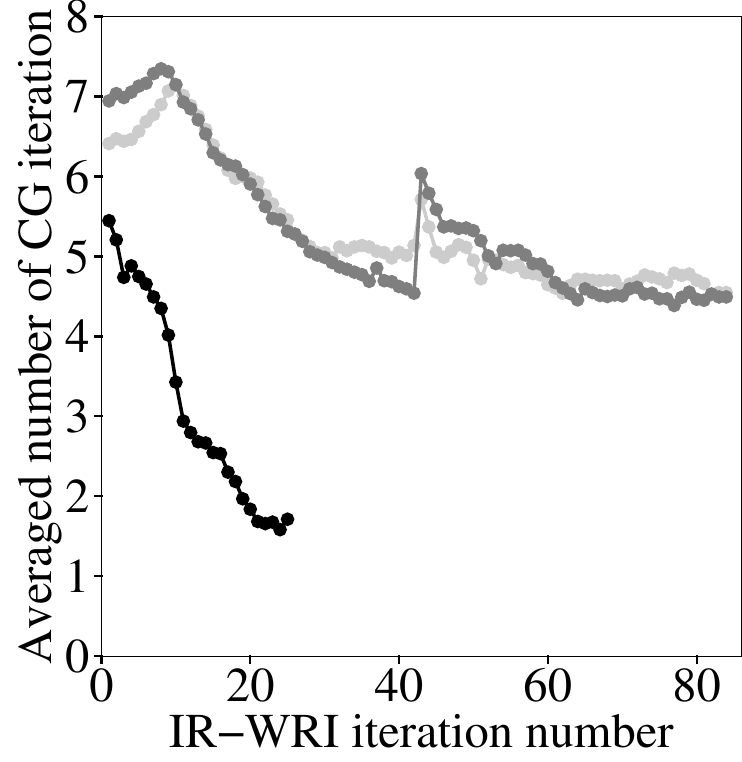}
\caption{Number of CG iterations averaged over source against IR-WRI iterations.} 
\label{BP_CG_num}
\end{figure}

\begin{figure}[ht!]
\centering
\includegraphics[width=0.8\linewidth]{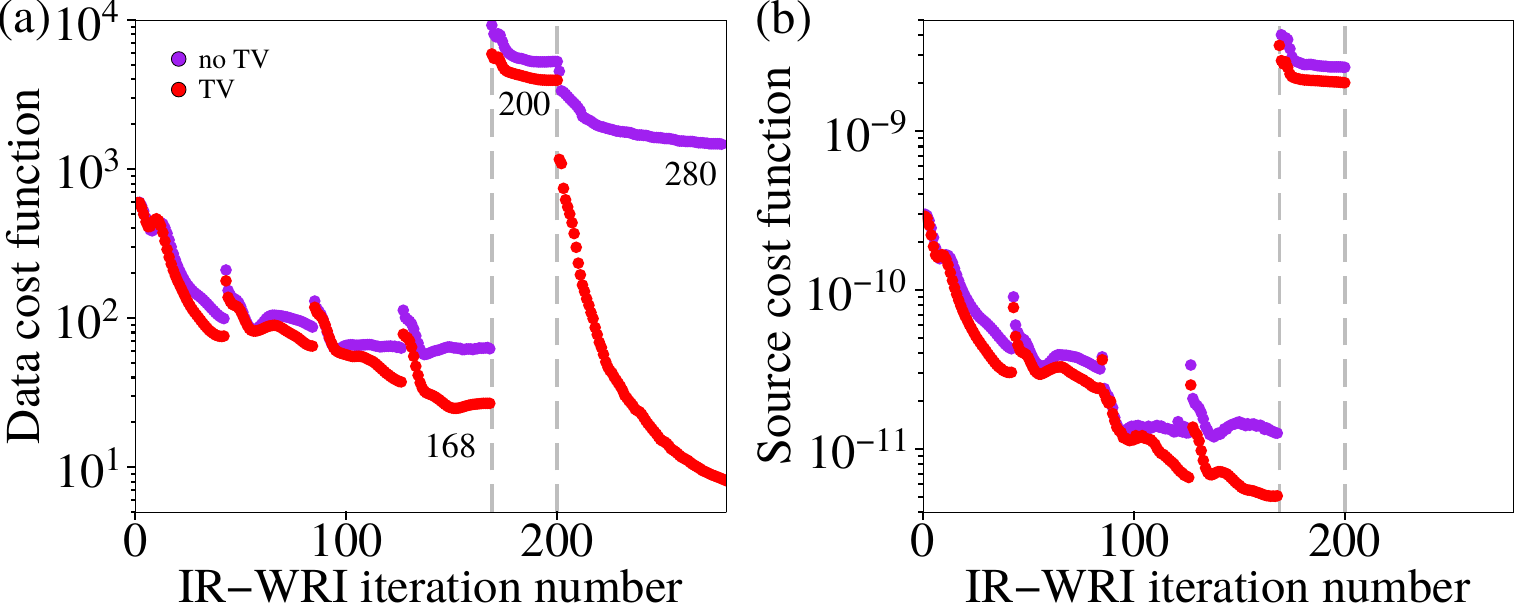}
\caption{(a) Data misfit function and (b) source  misfit function (b) against IR-WRI iterations for CG+2D-GMF methods. The dashed lines delineate the three multiscale steps. The number of IR-WRI/FWI iterations performed during each multiscale step are provided. Note that the first multi-scale step involves 84 IR-WRI iterations performed with CG method and 84 IR-WRI iterations performed with 2D-GMF method. Lagrange multipliers are restarted one time for the two series of IR-WRI iterations. Blue: Without TV regularization. Red: With TV regularization.} 
\label{BP_cost}
\end{figure}

\begin{figure}[ht!]
\centering
\includegraphics[width=0.7\linewidth]{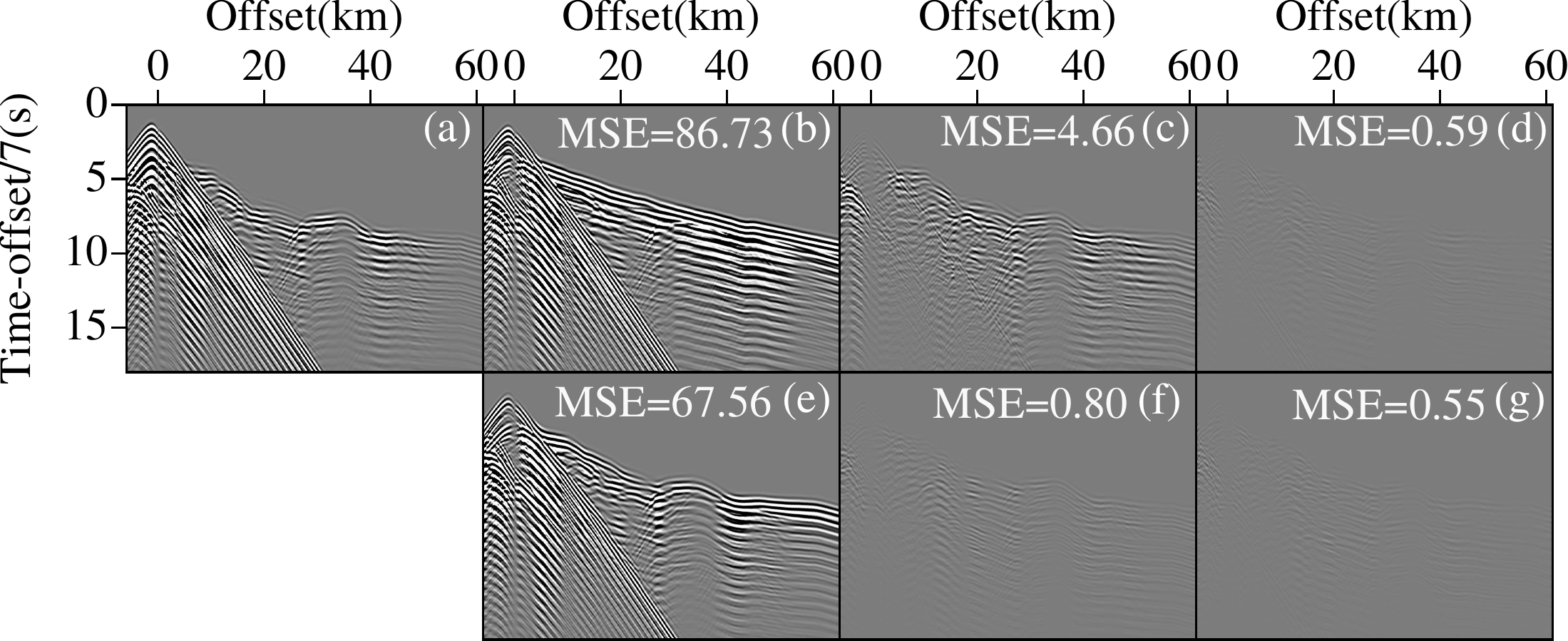}
\caption{Data fit in final FWI models. (a) Recorded data. (b) Data residuals for simulated data in starting model of Figure~\ref{BP_true_init}b. (c-d) Data residuals for simulated data in final FWI models of Figures~\ref{BP_inv}g and \ref{BP_inv_reg}g (without and with TV regularization during the first two multiscale steps, respectively). (e) Data residuals for simulated data in FASTT model (Figure~\ref{BP_inv_tomo}a). (f-g) Data residuals for simulated data in final FWI models of Figures~\ref{BP_inv_tomo}d,g (inferred from  SF and 2D-GMF+CG methods, respectively).} 
\label{BP_data_final}
\end{figure}

The first series of tests is performed without TV regularization. Note that, compared to the Marmousi test, we perform the first multi-scale step of Test 5 (Table~\ref{tab1}) by performing a certain number of IR-WRI iterations with the CG method before switching to the 2D-GMF method. Moreover, we periodically reinitialize the Lagrange multipliers during the IR-WRI iterations to speed-up the convergence of the alternating-direction method of multipliers \citep{Goldstein_2014_FAD}. The reconstructed velocity models during the first multiscale step are shown in Figure~\ref{BP_inv}a-e. The CG method is implemented with a stopping criterion of iteration defined by $\epsilon_1=0.1$ and $\epsilon_2=0.05$. With this criterion, the maximum and minimum numbers of CG iterations averaged over sources are around 7.1 and 4.5, respectively, during the first multiscale step (Figure~\ref{BP_CG_num}, light gray curve).
The SF and 1D matching filter methods manage to capture  the top of the salt, but they fail to reconstruct reliable deep structures where more inaccurate extended wavefields propagate (Figure~\ref{BP_inv}a-c). The reconstructed models are improved with the 2D-GMF and CG methods (Figure~\ref{BP_inv}d-e). However, the accuracy of the reconstructed model by the CG algorithm  remains insufficient to drive the second and the third multiscale steps toward a good velocity model (Figure~\ref{BP_inv}f-g). As shown in Figure~\ref{BP_cost}, blue curves, the cost function of the third multiscale step doesn't decrease sufficiently showing that FWI remains trapped in a local minimum. Note that the discontinuities of the data misfit function in Figure~\ref{BP_cost}a show the restart of the Lagrange multipliers. Moreover, we perform 84 additional IR-WRI iterations with the 2D-GMF method after the first 84 IR-WRI iterations with the CG method. The recorded data and the differences with the simulated data in the starting model and in the final FWI model (Figure~\ref{BP_inv}g) are shown in Figure~\ref{BP_data_final}(a-c). Although most of the kinematic mismatches have been solved by IR-WRI/FWI, the amplitude residuals between the recorded data and the simulated data in the FWI model still need to be further decreased.
%
%
%
%
%
%
%
%
%
%
%
%

\newpage
\clearpage

\subsubsection*{The effect of TV regularization on extended-source FWI}

\begin{figure}[ht!]
\centering
\includegraphics[width=0.6\linewidth]{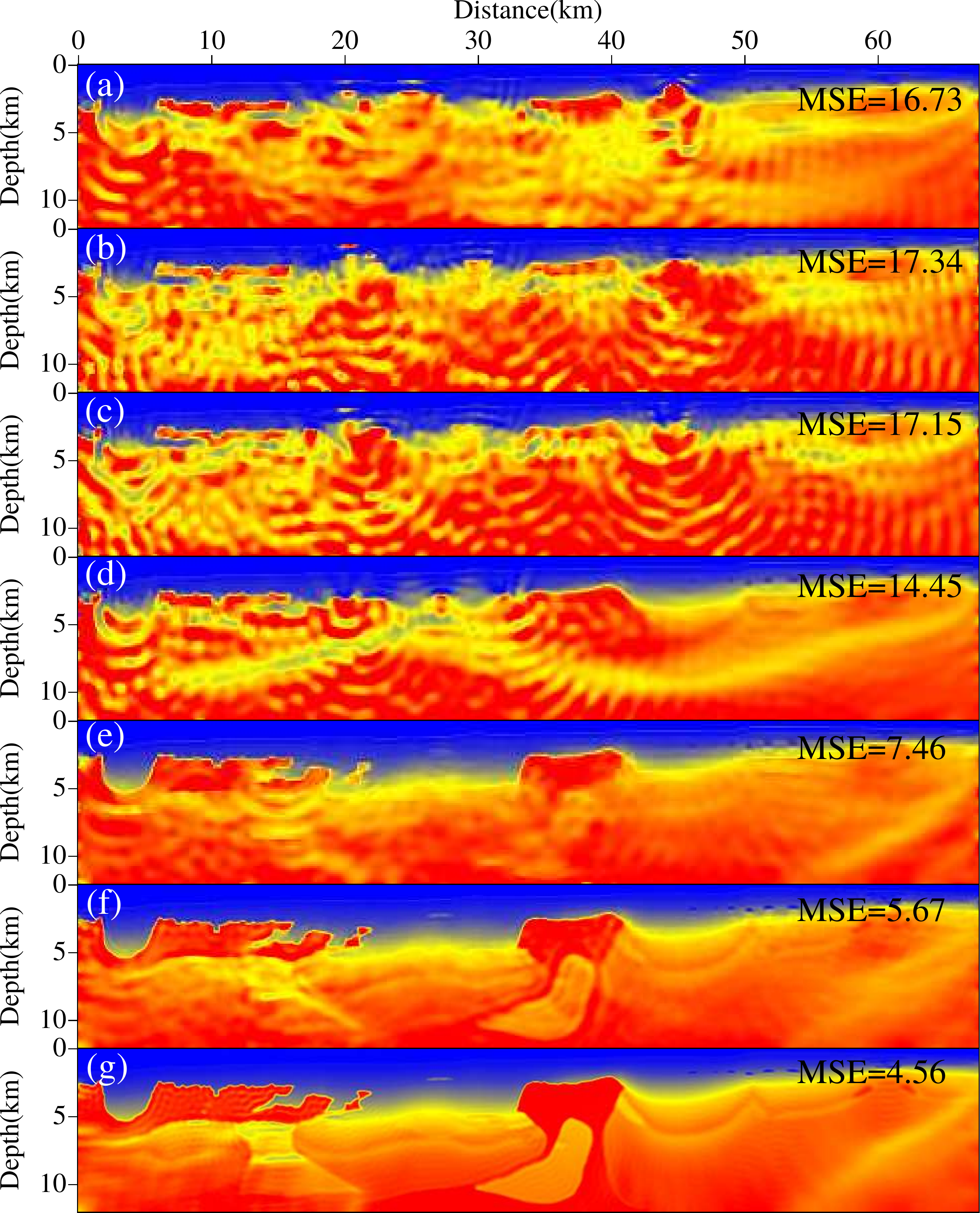}
\caption{Same as Figure~\ref{BP_inv} when TV regularization is used.} 
\label{BP_inv_reg}
\end{figure}

\begin{figure}[ht!]
\centering
\includegraphics[width=0.7\linewidth]{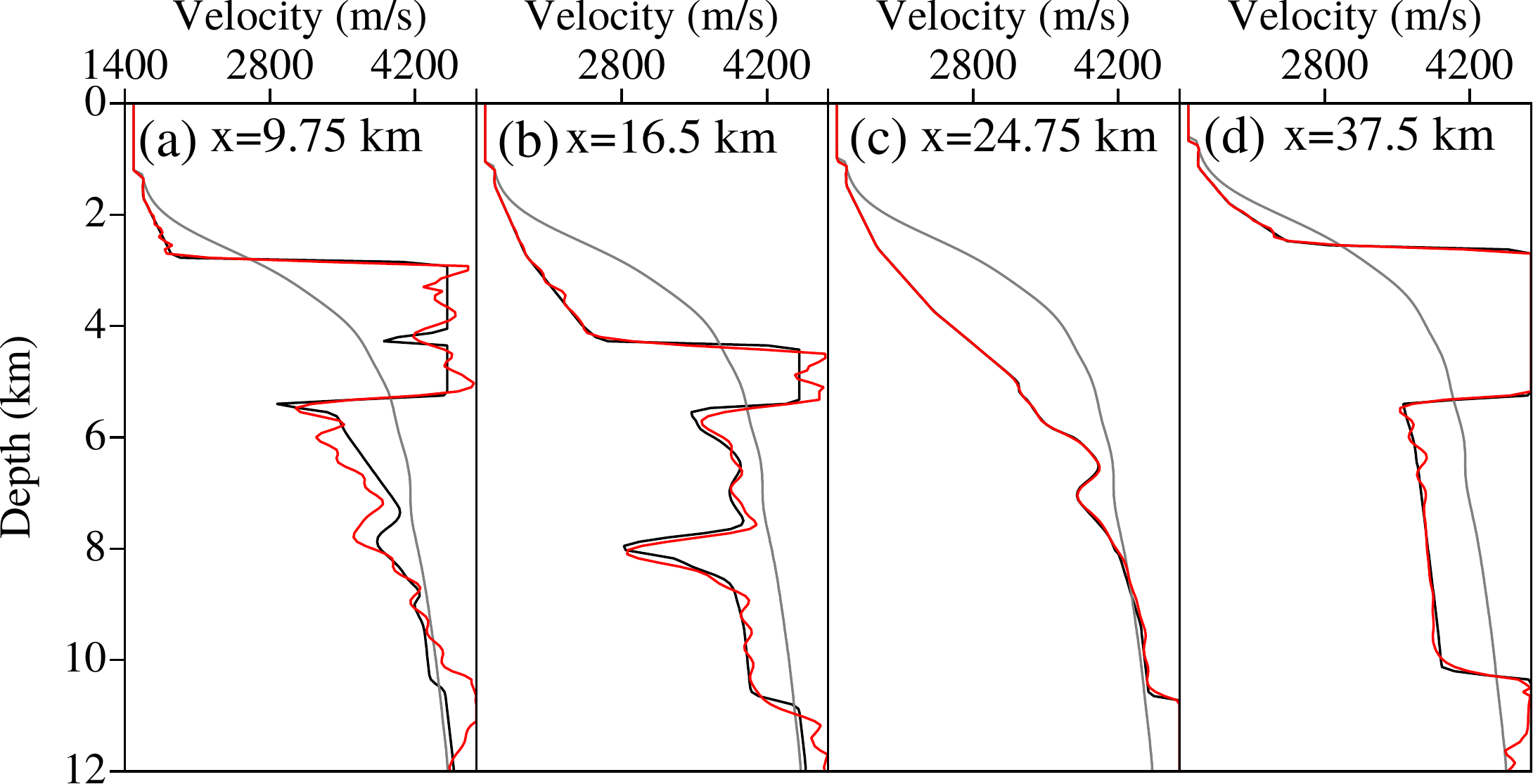}
\caption{Direct comparison between true model (black), initial model (gray) and reconstructed models with TV regularization and CG+2D-GMF (red) at (a) x=9.75~km, (b) 16.5~km, (c) 24.75~km and (d) 37.5~km.} 
\label{BP_logs_reg}
\end{figure}

\noindent We assess now the improvement of the inversion results provided by TV regularization (Figure~\ref{BP_inv_reg}). As expected, the TV regularization helps to better reconstruct the salt bodies while filtering out strong artifacts in the deep part of the model shown in Figure~\ref{BP_inv}. The estimated velocity models during  the first multiscale step are gradually improved as the data-domain Hessian is taken into account more accurately (Figure~\ref{BP_inv_reg}(a-e)). As in the previous test, we perform 84 IR-WRI iterations with the CG method before performing 84 iterations with the 2D-GMF method during the first multiscale step (Figure~\ref{BP_cost}, red curves). The number of CG iterations against the number of IR-WRI iteration during the first multiscale step is shown in Figure~\ref{BP_CG_num}, dark gray curve.  We use the final model of the first multiscale step inferred from the CG+2D-GMF method (Figure~\ref{BP_inv_reg}e) to perform the second and third multiscale steps. The final velocity models of these two steps are shown in Figure~\ref{BP_inv_reg}(f-g). The TV-regularized inversion converges toward a quite accurate velocity model, as further supported by the direct comparison between the true model, the initial model, and the final FWI model along vertical profiles (Figure~\ref{BP_logs_reg}). The convergence toward an accurate model is also illustrated by the convergence curves of the data and source misfit functions (Figure~\ref{BP_cost}, red curves), and by the data fit that has been dramatically improved compared to the case where no TV regularization is used (Figure~\ref{BP_data_final}d).
%
%
%
%

\newpage
\clearpage

\subsubsection*{On the sensitivity of extended-source FWI to the starting model}

\begin{figure}[ht!]
\centering
\includegraphics[width=0.6\linewidth]{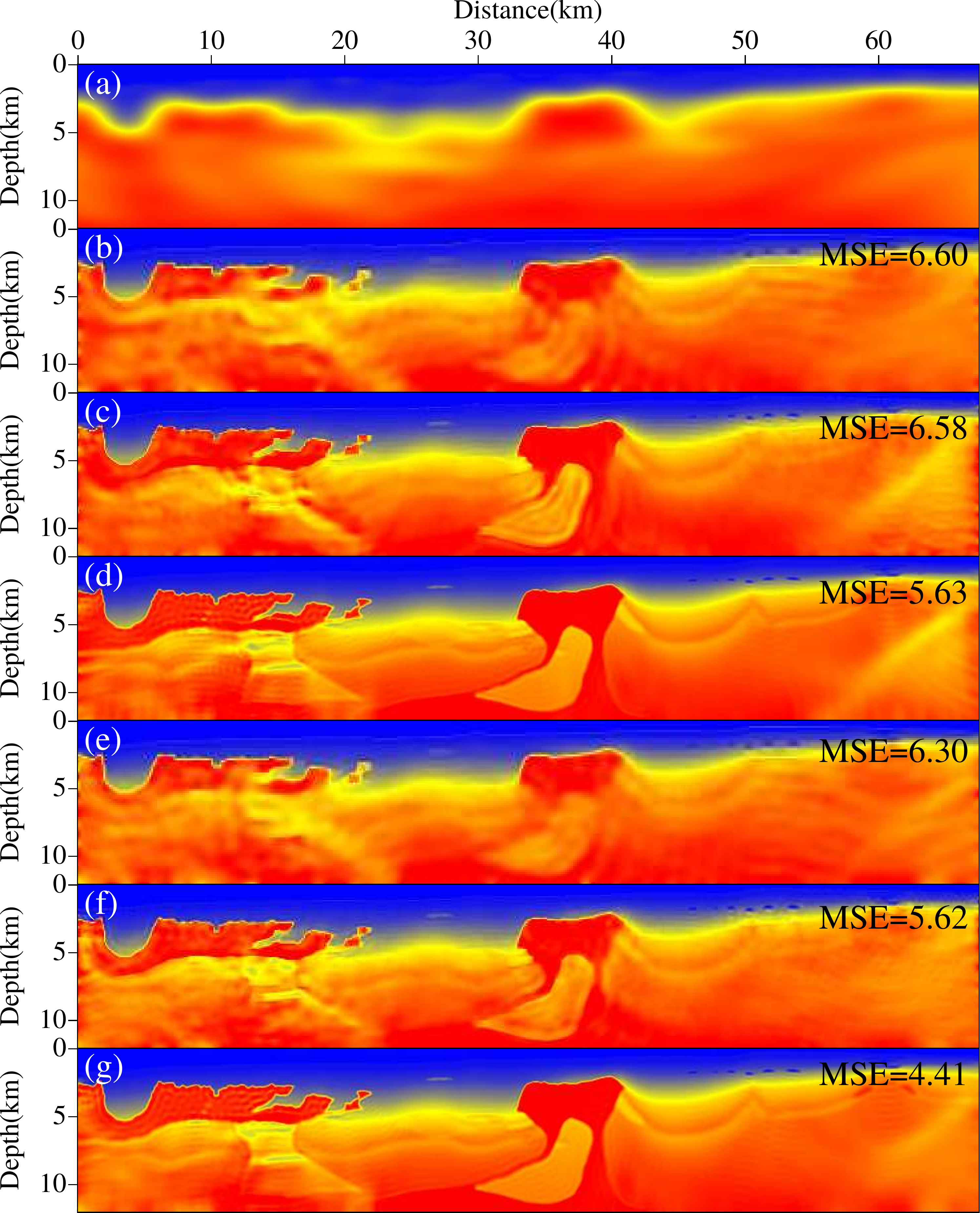}
\caption{(a) Initial model obtained by slope tomography \citep{Sambolian_2020_STF}. (b-d) Velocity models obtained by IR-WRI/FWI workflow performed with the SF method and TV regularization. (b-c) First (b) and second (c) multiscale steps performed with IR-WRI. (d) Third multiscale step performed by FWI. (e-g) Same as (b-d) when IR-WRI/FWI workflow performed with the 2D-GMF+CG method and TV regularization.} 
\label{BP_inv_tomo}
\end{figure}

\begin{figure}[ht!]
\centering
\includegraphics[width=0.8\linewidth]{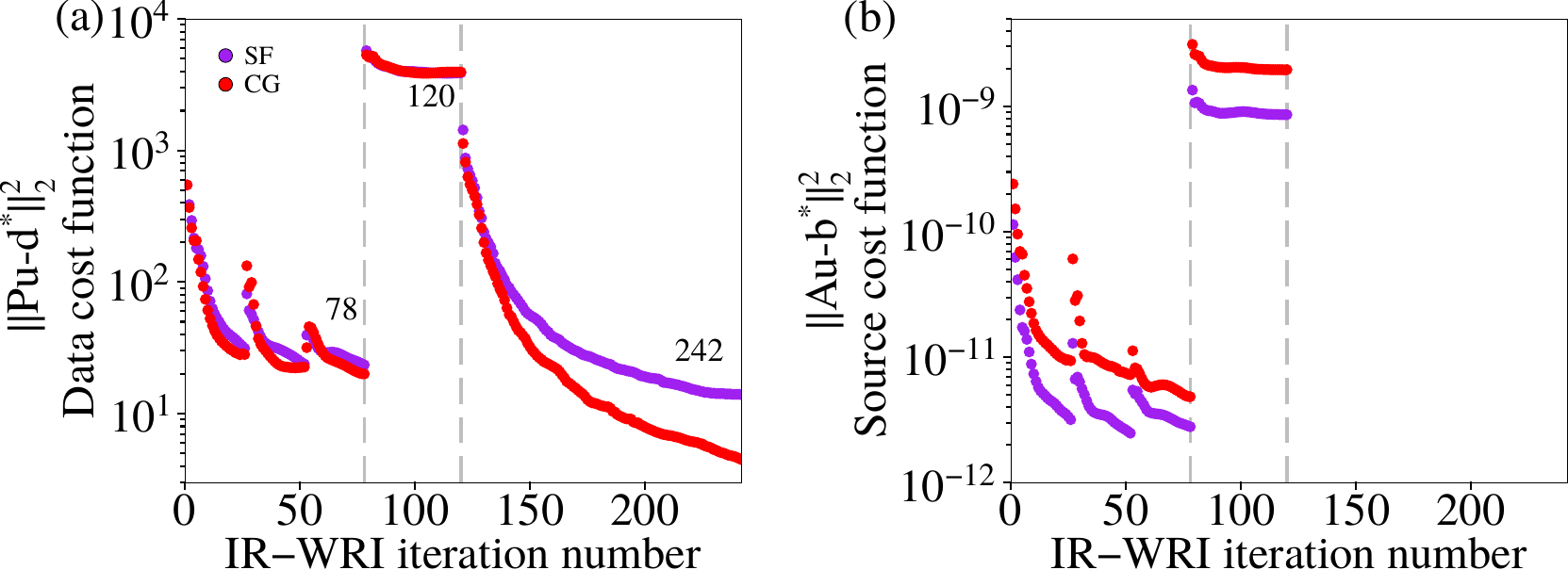}
\caption{(a) Data misfit function and (b) source misfit function against IR-WRI/FWI iterations when SF (purple line) and CG+2D-GMF (red line) methods are used during the first multiscale step. Note the smaller number of iterations compared to Figure~\ref{BP_cost}.} 
\label{BP_cost_tomo}
\end{figure}

\noindent A more accurate initial model allows for more accurate data-assimilated wavefield reconstruction (namely, wavefields that are closer to the true wavefields). Then, these more accurate wavefields should lead to a faster convergence of IR-WRI and relax the need to account for the effect of the data-domain Hessian accurately. Put simply, beginning IR-WRI from more accurate initial models is another leverage to mitigate its computational burden. To illustrate this statement, we perform IR-WRI with an initial velocity model built from a long-offset dataset by first-arrival traveltime+slope tomography (FASTT) \citep{Sambolian_2020_STF} (Figure~\ref{BP_inv_tomo}a). We perform the multiscale IR-WRI/FWI workflow with the two end-members SF and 2D-GMF+CG methods and we apply TV regularization. As shown in Figure~\ref{BP_inv_tomo}(b-d), the IR-WRI/FWI workflow performed with the SF method converges toward an acceptable model due to the more accurate initial model with however small artifacts in the subsalt area. The multiscale inversion performed with the 2D-GMF+CG method further improves significantly the resolution and the accuracy of the subsalt imaging at reservoir depths (Figure~\ref{BP_inv_tomo}(e-g)). Moreover, the accurate initial model dramatically decreases the number of CG iterations averaged over sources performed during each IR-WRI iteration (Figure~\ref{BP_CG_num}, black curve) as well as the convergence of the data misfit and source misfit functions (compare Figure~\ref{BP_cost_tomo} and Figure~\ref{BP_cost}). The data residuals between the recorded data and the simulated data in the starting FASTT model and the final FWI models inferred from the SF and 2D-GMF+CG methods (Figures~\ref{BP_inv_tomo}d,g) are also shown in Figure~\ref{BP_data_final}(e-g) and can be compared with those obtained when the inversion starts from the crude initial model Figure~\ref{BP_data_final}(b-d).
%
%
%
%

\newpage
\clearpage

%
\section{Discussion}

\noindent In this study, we design a practical workflow to implement extended-source FWI in the time domain. In extended-source FWI, the wave equation is solved with data-driven source extensions to compute wavefields that are closer to the true wavefields. Then, these improved wavefields are used to compute more accurate virtual sources in the sensitivity kernel of FWI beyond the Born approximation (see \citet{Operto_2022_FBA} for a tutorial review). The source extensions are the damped least-squares solution of the scattered data fitting problem, where the recorded scattered data are the FWI data residuals of the current iteration. Accordingly, the source extensions are the solution of an adjoint-state wave equation, whose sources  are the weighted data residuals by the data-domain Hessian of the scattered-data fitting problem. As such, the source extensions play also the role of the so-called adjoint wavefields (i.e., scattering sources) in classical FWI. Accounting for this data-domain Hessian is the main computational bottleneck of extended-source FWI. The key contributions of this study are first to propose different strategies to account approximately for the effect of the data-domain Hessian with matching filters and truncated Gauss-Newton method and second to propose a practical workflow that progressively relaxes the accuracy with which the data-domain Hessian is taken into account while converging toward accurate subsurface models.

\subsection{Preconditioner for CG method and direct solver}
\noindent The CG method is a good candidate to solve approximately (Gauss-)Newton system as shown in the framework of FWI \citep{Metivier_2013_TRU}, extended-space FWI \citep{Fu_2017_AMA} or least-squares reverse time migration \citep{Hou_2016_AEL,   
Guitton_2017_FLR}.
However, it has two shortcoming in the framework of extended-source FWI: First, when a crude initial model is used and the free surface complicates the data anatomy, the convergence of the CG method becomes extremely slow due to the poor condition number of the data-domain Hessian. Second, every iteration of the CG method involves two wavefield simulations (by source), and this computational overhead may hinder applications on large-scale 3D problems. In this study, we use approximated inverse of the  data-domain Hessian to build approximate solutions of the Gauss-Newton system. We use these approximate solutions in the place of the exact solutions or as starting points of an iterative refinement procedure implemented with the CG method. We have shown that these approximate solutions may not be accurate enough to converge toward acceptable velocity model when dealing with complex media such as the BP salt model and crude initial models. Moreover, the number of CG iterations to reach accurate solutions of the Gauss-Newton system may be prohibitively large without preconditioner even when the surrogate solutions are used as starting guess. Therefore, designing a good preconditioner of the Gauss-Newton system would speed up significantly the inversion and further broaden the range of applicability of extended-source FWI. A first possible option is to build a preconditioner with analytical Green functions in homogeneous or velocity gradient models. A second option would be to build a preconditioner in the ray-theoretical framework considering that this preconditioner is mostly needed during the first steps of the multiscale inversion (i.e., at low frequencies) when the subsurface models are smooth (i.e., when ray theory is valid). A third option would be to compute the data-domain Hessians explicitly in the frequency domain for each frequency, solve the normal systems exactly for each frequency with a direct solver before transforming back the deconvolved data residuals in the time domain by inverse Fourier transform. Alternatively, a diagonal preconditioner for the CG method could be computed with this approach to avoid using direct solvers. Finally, the matching filters estimated in this study may be used as preconditioner rather than as starting guess by the CG method (see \citet{Guitton_2017_FLR} for an application in least-squares reverse-time migration). Designing the best strategy among the proposed approaches will be the aim of future works.

\subsection{Free-surface multiples and layer stripping approaches}

\noindent A second issue is related to the convergence of extended-source FWI in the case of complex models and complex data (i.e., with multiples). \citet{Operto_2022_FBA} reviewed that the data-assimilated wavefields are reconstructed by a migration/demigration of the recorded data. Accordingly, their accuracy decreases at they are propagated backward in time in the background model away from the receivers during the migration step. This inaccuracy of the reconstructed wavefields away from the receivers can trap the inversion in spurious minima and this nonlinearity is clearly exacerbated  when free surface multiples are involved in the inversion. Accounting for the data-domain Hessian with good preconditioner or direct solvers may be an option to deal with complex datasets. Alternatively, data-driven continuation strategies such as layer stripping may be the recipe to feed progressively the inversion with more complex data and further extend its linear regime accordingly. They can be implemented with appropriate covariance matrices or annihilators in the data misfit function under the form of time and offset windowing (see \citet{Gorszczyk_2017_TRW} in the framework of classical FWI) and/or in the source misfit function \citep{Lee_2020_SFW,Huang_2018_SEW,Rizzuti_2021_DFW}. The design of the optimal covariance matrices will be the second focus of future studies.

\subsection{On the tuning of extended-source FWI}
We have shown how the accuracy with which the data-domain Hessian is taken into account can be progressively relaxed during multiscale inversion. Typically, the truncated Gauss-Newton method may be only necessary during the first multiscale steps at low frequencies or may be disregarded in the case of sufficiently accurate starting model. During intermediate multiscale steps, 2D Gabor matching filter should provide a sufficiently accurate estimate of the data-domain inverse to avoid the CG iterations. Finally, one can switch to FWI during the latest multiscale steps. The optimal design may be quite case dependent and our workflow needs now to be assessed against real data applications. However, switching from IR-WRI to FWI is easy from the implementation viewpoint since it amounts to disable the source extensions in the right-hand side of the wave equation during wavefield reconstruction, replace the data-domain Hessian by the identity matrix in the adjoint source and disregard the Lagrange multipliers \citep{Operto_2022_FBA}.

%
\section{Conclusions}

\noindent The bottleneck of extended-source FWI in the time domain is the solution of the least-squares scattered-data fitting problem for source extension reconstruction. This can be implemented by time-reversed modeling of the source-dependent data residuals deblurred by the inverse of the data-domain Hessian. We propose a practical workflow based on 1D/2D matching filters in the Fourier and short-time Fourier domains and the conjugate-gradient method to account for this Hessian with variable accuracy and find the best compromise between imaging accuracy and computational cost. This workflow has already allowed us to image a complex model such as the 2004 BP salt model from long-offset data, realistic starting frequency (1.5~Hz) and a crude starting model.
We first show that that the data-domain Hessian allows for a more accurate amplitude match of the data during wavefield reconstruction, which is critical during the early steps of extended-source FWI. Sparsity-promoting TV regularization is another key ingredient to further mitigate noise during imaging and deal with complex media with sharp contrasts. We also show that more complex datasets with free-surface multiples require more accurate estimation of the data-domain Hessian. Conversely, building more accurate starting velocity models with tomography methods is an efficient leverage to mitigate the need of accurate data-domain Hessian estimation. Future works aim first at designing efficient preconditioner of the Gauss-Newton system to minimize the computational overhead of the CG method and broaden the applicability of extended-source FWI accordingly and second at designing covariance matrices in the data and source misfit functions to further extend the linear regime of extended-source FWI with layer-stripping strategies. Finally, understanding the potential and limits of the proposed workflow against real data case studies will help to understand how to systematically design the optimal workflow of extended-source FWI depending on the specifications of the application.

\noindent \textbf{Acknowledgments:} This study was funded by the WIND consortium (https://www.geoazur.fr/WIND), sponsored by Chevron, Petrobras, Shell, and Total (2020-2022) and AkerBP, ExxonMobil, Petrobras, Shell and SINOPEC (2023-2025). The authors are grateful to the OPAL infrastructure from Observatoire de la Côte d’Azur (CRIMSON) for providing resources and support. This work was granted access to the HPC resources of IDRIS under the allocation A50050410596 made by GENCI. Gaoshan Guo would like to thank Youshan Liu (IGGCAS), Qiancheng Liu (IGGCAS) and Peng Yong (ISTerre) for fruitful discussions about matching filters. We thank Serge Sambolian (ISTerre-UGA) for providing us the FASST model and for suggesting the idea of matching filters for Hessian estimation.

\append{The adjoint-state method for the first-order velocity-stress wave equation}

\noindent We review how to compute the gradienf of FWI with the adjoint-state method when the wave equation is solved with a velocity-stress formulation, equation~\ref{1steq}. In the framework of the adjoint-state method \citep{Plessix_2006_RAS}, the FWI misfit function can be recast as a Lagrangian function as 
\begin{equation}
\begin{aligned}
\mathcal{L}(\mathbf{u}(\mathbf{x},t),\mathbf{v}(\mathbf{x},t),\mathbf{m}(\mathbf{x})) &= \frac{1}{2} \int_0^T \sum_r (\mathbf{Pu}(\mathbf{x},t)-\mathbf{d}^*(t))^2  dt \\
& + \int_0^T \left\langle \mathbf{v}(\mathbf{x},t), \mathbf{C(m)} \partial_t \mathbf{u}(\mathbf{x},t) - \mathbf{Du}(\mathbf{x},t) - \mathbf{b}^*(t) \right\rangle dt,
\end{aligned}
\end{equation}
where $\mathbf{v}$ is Lagrangian multiplier or adjoint variable and the wave-equation constraint is written in a self-adjoint form, equation~\ref{wave}. Using integration by part for time and space derivatives, the new Lagrangian reads
\begin{equation}
\begin{aligned}
\mathcal{L}(\mathbf{u}(\mathbf{x},t),\mathbf{v}(\mathbf{x},t),\mathbf{m}(\mathbf{x})) & = \frac{1}{2} \int_0^T \sum_r (\mathbf{Pu}(\mathbf{x},t) -\mathbf{d}^*(t))^2  dt + \left\langle \mathbf{u}(\mathbf{x},T), \mathbf{C(m)} \mathbf{v}(\mathbf{x},T) \right\rangle \\ 
& - \int_0^T \left\langle \mathbf{u}(\mathbf{x},t), \mathbf{C(m)} \partial_t \mathbf{v}(\mathbf{x},t) \right\rangle dt
+ \int_0^T \left\langle \mathbf{u}(\mathbf{x},t), \mathbf{Dv}(\mathbf{x},t)  \right\rangle dt \\
& - \int_0^T \left\langle \mathbf{v}(\mathbf{x},t), \mathbf{b} \right\rangle dt.
\end{aligned}
\end{equation} \\
Minimizing the cost function with respect to state variable $\mathbf{u}$ gives the adjoint-state equation
\begin{equation}
\mathbf{C(m)} \partial_t \mathbf{v}(\mathbf{x},t) = \mathbf{Dv}(\mathbf{x},t) + \sum_r \mathbf{P}^T (\mathbf{Pu}(\mathbf{x},t)-\mathbf{d}^*(t) ),
\end{equation}
or, equivalently in a form suitable for explicit time stepping
\begin{equation}
\partial_t \mathbf{v}(\mathbf{x},t) = \mathbf{M(m)}  \mathbf{Dv}(\mathbf{x},t) + \mathbf{M(m)}  \sum_r \mathbf{P}^T ( \mathbf{Pu}(\mathbf{x},t) - \mathbf{d}^*(t)),
\end{equation}
with a final condition $\mathbf{v}(\mathbf{x},T)=0$. The final condition is transformed to initial condition with a change of variables,
\begin{equation}
\partial_t \mathbf{w}(\mathbf{x},t) = \mathbf{M(m)}  \mathbf{Dw}(\mathbf{x},t) + \mathbf{M(m)}  \sum_r \mathbf{P}^T ( \mathbf{Pu}(\mathbf{x},t) - \mathbf{d}^*(t))
\end{equation}
where $\mathbf{w}(\mathbf{x},t)=\mathbf{v}(\mathbf{x},T-t)$ and $\mathbf{w}(\mathbf{x},0)=0$. Then, we reverse the time variable by $t'=T-t$
\begin{equation}
\partial_t \mathbf{w}(\mathbf{x},t') = -\mathbf{M(m)}  \mathbf{Dw}(\mathbf{x},t') - \mathbf{M(m)} \mathbf{P}^T \sum_r ( \mathbf{Pu}(\mathbf{x},T-t') - \mathbf{d}^*(T-t'))
\end{equation}
with initial condition $\mathbf{w}(x,0)=0$. The same code can be used to solve the state and adjoint-state equations. In the later case, one needs to use negative time step and reverse the data residuals in time.


\append{Implementation of TV regularization in extended-source FWI}

\noindent In this study, we consider a non-smooth TV regularization, a mixed $l_2$ and $l_1$ norm (for more details, see \citet{ Aghamiry_2019_IBC,Aghamiry_2019_CRO}), 
\begin{equation}
\mathcal{R}(\mathbf{m}) = \| \mathbf{m} \|_{TV} = \sum \| \left( \nabla_x \mathbf{m} ~ \nabla_z \mathbf{m} \right) \| = \sum \sqrt{ \left( \nabla_x \mathbf{m} \right) ^2 + \left( \nabla_z \mathbf{m} \right)^2},
\end{equation}
where $(\mathbf{p}_1 ~ \mathbf{p}_2)$ denotes a two-column matrix, $\| (\mathbf{p}_1 ~ \mathbf{p}_2) \| = \sqrt{\mathbf{p}_1^2 + \mathbf{p}_2^2}$ is a vector which contains the $\ell{2}$ norm of each row of $(\mathbf{p}_1 ~ \mathbf{p}_2)$ and $\sum \sqrt{\mathbf{p}_1^2 + \mathbf{p}_2^2}$ is the $\ell{1}$ norm of $\| (\mathbf{p}_1 ~ \mathbf{p}_2) \|$.
We aim at solving the following regularized problem for subsurface parameters $\mathbf{m}$
\begin{equation}
\min_{\mathbf{m}} \mathcal{P}_{\mu}(\mathbf{m}) + \lambda \mathcal{R}(\mathbf{m}).
\end{equation}
where  $\mathcal{P}_{\mu}(\mathbf{m})$ is the augmented Lagrangian function given in equation~\ref{eqwri} (the wavefields are assumed known and are kept fixed in the frame of the alternating direction method of multipliers (ADMM).

\noindent We introduce the auxiliary variable $\mathbf{p}$ to decouple the $\ell{1}$ problem and the $\ell{2}$ problem \citep{Goldstein_2009_SBM} and recast the former as a denoising problem tackled wih proximal algorithms \citep{Combettes_2011_PRO},
\begin{equation}
\min_{\mathbf{m}, \mathbf{p}} \mathcal{P}_{\mu}(\mathbf{m}) + \lambda \| \mathbf{p} \|_1 ~~ \text{subject to} ~~ \nabla \mathbf{m} = \mathbf{p},
\end{equation}
where
\begin{equation}
\mathbf{p}=\left( \nabla_x \mathbf{m} ~ \nabla_z \mathbf{m} \right).
\end{equation}

\noindent Solving this problem with an augmented Lagrangian method leads to a saddle point problem
\begin{equation}
\min_{\mathbf{m},\mathbf{p}} \max_{\mathbf{v}} \mathbb{P}_{\mu_k}(\mathbf{m}) + \lambda \| \mathbf{p} \|_1 - \left\langle \mathbf{v}, \mathbf{p} - \nabla \mathbf{m} \right\rangle + \frac{1}{2c} \| \mathbf{p} - \nabla \mathbf{m} \|^2_2.
\end{equation}
where $c$ is a penelay parameter. 

\noindent The primal and dual variables are updated in alternating mode in the framework of the method of multipliers,
\begin{subequations}
\begin{align}
(\mathbf{m}_{k+1}, \mathbf{p}_{k+1}) &= \argmin_{\mathbf{m}, \mathbf{p}} \mathbb{P}_{\mu_k}(\mathbf{m}) + \lambda \| \mathbf{p} \|_1 - \left\langle \mathbf{v}, \mathbf{p} - \nabla \mathbf{m}  \right\rangle + \frac{1}{2c} \| \mathbf{p} - \nabla \mathbf{m} \|^2_2, \\
\mathbf{v}_{k+1} &= \mathbf{v}_{k} - \frac{1}{c} ( \mathbf{p}_{k+1} - \nabla \mathbf{m}_{k+1}).
\end{align}
\end{subequations}

\noindent Using scaled Lagrange multipliers by penalty parameters, $\mathbf{q}_k= - c \mathbf{v}_k$, leads to the scaled form of the method of multipliers  \citep[][ Section 3.1.1]{Boyd_2011_DOS}
\begin{subequations}
\begin{align}
(\mathbf{m}_{k+1}, \mathbf{p}_{k+1}) &= \argmin_{\mathbf{m}, \mathbf{p}} \mathbb{P}_{\mu_k}(\mathbf{m}) + \lambda \| \mathbf{p} \|_1 + \frac{1}{2c} \| \mathbf{p} + \mathbf{q}_k - \nabla \mathbf{m} \|^2_2, \\
\mathbf{q}_{k+1} &= \mathbf{q}_{k} - ( \nabla \mathbf{m}_{k+1} - \mathbf{p}_{k+1} ),
\end{align}
\end{subequations}
Finally, primal variables are updated in alternative mode in the framework of ADMM
\begin{subequations}
\begin{align}
\mathbf{m}_{k+1} &= \argmin_{\mathbf{m}} \mathbb{P}_{\mu_k}(\mathbf{m}) + \frac{1}{2c} \| \mathbf{p}_k + \mathbf{q}_k - \nabla \mathbf{m} \|^2_2, \\
\mathbf{p}_{k+1} &= \argmin_{\mathbf{p}} \lambda \| \mathbf{p} \|_1 + \frac{1}{2c} \| \mathbf{p} - ( \nabla \mathbf{m}_{k+1} - \mathbf{q}_k) \|^2_2, \\
\mathbf{q}_{k+1} &= \mathbf{q}_{k} - ( \nabla \mathbf{m}_{k+1} - \mathbf{p}_{k+1} ).
\end{align}
\end{subequations}

\noindent Model parameters are updated according to 
\begin{equation}
\mathbf{m}_{k+1} = \mathbf{m}_k - \left( \mathbf{H}_k + \frac{1}{c} \nabla^T \nabla \right)^{-1}  \left( \mathbf{g}_k + \frac{1}{c} \nabla^T \left( \mathbf{p}_k + \mathbf{q}_k - \nabla \mathbf{m}_k \right) \right)
\end{equation}
where $\mathbf{H}_k$ and $\mathbf{g}_k$ are the Hessian and the gradient of $\mathbb{P}_{\mu_k}(\mathbf{m})$ for $\mathbf{m}=\mathbf{m}_k$.

\noindent The penalty parameter $c$ can be determined by a dynamic method as
\begin{equation}
c_k = \beta_k \frac{\| \nabla^T \left( \mathbf{p}_k + \mathbf{q}_k - \nabla \mathbf{m}_k \right) \|_2^2}{\| \mathbf{g}_k \|_2^2},
\end{equation}
where $\beta$ is a dimensionless scaling factor estimated by trial and error. Furthermore, it is typically decreased from 0.3 to 0.05 in iterations to reduce the weight of TV regularization in the objective function.

\noindent The $\mathbf{p}$ subproblem is a $l_1$-norm regularization denoising problem
\begin{equation}
\mathbf{p}_{k+1} = \text{prox}_{\lambda c \| \mathbf{p} \|_1} (\nabla \mathbf{m}_{k+1} - \mathbf{q}_k) = \argmin_{\mathbf{p}} \| \mathbf{p} \|_1 + \frac{1}{2 \lambda c} \| \mathbf{p} - ( \nabla \mathbf{m}_{k+1} - \mathbf{q}_k) \|^2_2,
\end{equation}

\noindent The solution can be written as
\begin{equation}
\mathbf{p}_{k+1}=\text{sign}(\mathbf{z})\max(\| \mathbf{z} \| - \lambda c, 0),
\end{equation}
where $\mathbf{z}=\nabla \mathbf{m}_{k+1} - \mathbf{q}_k$. The optimal $\lambda c=\alpha \max(\| \mathbf{z} \|)$, where $\alpha$ controls the soft thresholding performed by the TV regularization. It is set to 0.2 in this study.

\bibliographystyle{seg}
\bibliography{windbiblio,windtemp}

\end{document}